\documentclass[fleqn,usenatbib]{mnras}

\usepackage{newtxtext,newtxmath}

\usepackage[T1]{fontenc}

\DeclareRobustCommand{\VAN}[3]{#2}
\let\VANthebibliography\thebibliography
\def\thebibliography{\DeclareRobustCommand{\VAN}[3]{##3}\VANthebibliography}

\usepackage{graphicx}	
\usepackage{amsmath}	

\title[CHEFT]{CHEFT: A Hybrid Effective Field Theory halo  model}

\author[Pellejero Ibáñez et al.]{
Marcos Pellejero Ibáñez$^{1}$\thanks{E-mail: mpelleje@ed.ac.uk}, David Alonso$^{2}$, John A. Peacock$^{1}$, Matteo Zennaro$^{3}$, and Samuel Brieden$^{4}$
\\
\\
$^{1}$Institute for Astronomy, University of Edinburgh, Royal Observatory, Blackford Hill, Edinburgh, EH9 3HJ, UK\\
$^{2}$Department of Physics, University of Oxford, Denys Wilkinson Building, Keble Road, Oxford OX1 3RH, United Kingdom \\
$^{3}$Institute of Space Sciences (ICE, CSIC), Campus UAB, Carrer de Can Magrans, s/n, 08193 Barcelona, Spain \\
$^{4}$Institute for Theoretical Particle Physics and Cosmology (TTK), RWTH Aachen University, Sommerfeldstr. 16, D-52056 Aachen, Germany
}

\date{Accepted XXX. Received YYY; in original form ZZZ}

\pubyear{\the\year{}}

\begin{document}
\label{firstpage}
\pagerange{\pageref{firstpage}--\pageref{lastpage}}
\maketitle

\begin{abstract}
  We present a hybrid halo model, which improves the description of the 2-halo term by incorporating non-linear information from simulations. A linear computation of the halo-halo power spectrum is inaccurate at the transition between the 1-halo and 2-halo regimes, whereas nonlinear approaches such as Hybrid Effective Field Theory (HEFT) are not naturally compatible with the halo model decomposition. We address this limitation by constructing a collapsed HEFT (CHEFT) framework, in which the power-spectrum templates of the HEFT operator expansion are measured from simulations where 1-halo contributions are removed by collapsing particles to their halo centres. The halo-halo power spectrum is then expressed as a sum over bias operators, with mass-dependent bias parameters deduced from simulation using the probabilistic bias approach. This provides a predictive model in which there are no free bias parameters.
  We validate the model for a range of weighting schemes designed to mimic the halo-mass dependence of astrophysical observables, including the Sunyaev--Zeldovich effect, the Cosmic Infrared Background, and galaxy abundances described via a halo occupation distribution. For the matter field, the model recovers the power spectrum to percent-level accuracy across the transition regime. For weighted tracers, the baseline model achieves accuracies of $\sim$\,$5-10\%$ in power, which improves to the $\sim$\,$3-5\%$ level when including an effective higher-derivative, Laplacian-like contribution in the bias expansion. The CHEFT model thus retains the precision and flexibility of the EFT approach, while allowing the transparent incorporation of astrophysical effects that are directly associated with haloes.

\end{abstract}




\section{Introduction}
  The large-scale distribution of matter in the Universe encodes a wealth of cosmological information, making accurate theoretical predictions for clustering statistics across a broad range of large-scale structure (LSS) tracers a central goal of modern cosmology.

  The halo model \citep[e.g.][]{astro-ph/0001493,astro-ph/0005010} offers a physically motivated description of clustering across a wide range of scales, as well as a flexible formalism for describing the correlated structure of a great variety of astrophysical probes. Indeed, the halo model has been the framework of choice in the description of small-scale clustering of matter \citep[e.g.][]{1505.07833,2009.01858}; the abundance and clustering of galaxies \citep[e.g.][]{astro-ph/0109001,1206.6890,2405.13495}; the properties of the thermal gas pressure and density through the Sunyaev--Zeldovich effect
  \citep[e.g.][]{astro-ph/0205468,1109.3711,2005.00009,2507.07346}; the statistics of star formation as mapped by the Cosmic Infrared Background (CIB) 
  \citep[e.g.][]{1109.1522,2006.16329,2206.15394}; the X-ray emissivity 
  \citep[e.g.][]{1909.02179,2410.22397,2309.11129}; and many other probes 
  \citep[e.g.][]{1301.5901,1312.4403,2307.14881}.
  
  In its standard form, the halo model power spectrum is decomposed into 1-halo contributions from particles within the same halo and 2-halo terms from elements residing in different haloes. While this decomposition captures the qualitative behaviour of the matter power spectrum, its quantitative accuracy is limited by several sources of uncertainty, as reviewed by \citet{2303.08752}. At the top of this list is the approximate treatment of the 2-halo term, which requires an accurate description of the quasilinear clustering of haloes of different masses. In traditional implementations, halo clustering simply assumes a linear halo bias relation combined with the linear matter power spectrum. This typically fails to match the non-linear halo-halo power spectrum on intermediate scales, $k\sim 0.3$-$1\,h\,{\rm Mpc}^{-1}$ \citep{astro-ph/0411777,1406.5013,2011.08858}, degrading the description of the transition between the 1-halo and 2-halo-dominated regimes. 

  Intermediate scales can be tackled by effective field theory: this provides a systematic description of clustering, incorporating non-linear bias and higher-order contributions in a controlled perturbative expansion. (EFT: \citealt{1004.2488,1310.0464,1311.2168,1406.7843,1506.05264,1412.5169,1611.09787,2005.00523,2012.04636}). EFT can be fruitfully extended to Hybrid Effective Field Theory (HEFT, \citealt{1910.07097,2109.08699,2103.09820,2107.10287,2101.11014,2101.12187}), which expresses the density field of tracers as a combination of Lagrangian operators constructed from the initial conditions, combined with using exact $N$-body simulation displacements in order to map the initial particle grid into the complex non-linear Eulerian space at a given redshift, $z$. This weighting of particles by the Lagrangian operators has been shown to work to high accuracy on scales larger than the typical halo size ($k\sim0.6\,h{\rm Mpc}^{-1}$). However, tracers inside haloes exhibit either more complex dependencies on their Lagrangian environments than the simple polynomial expansion assumed by HEFT, or reduced dependencies after several orbits around the halo centre. A clean solution to this problem would be to model the 1-halo term separately; but the standard HEFT approach does not naturally separate 1-halo and 2-halo contributions, so that any combination of the HEFT and halo model approaches risks double counting of non-linearities on sub-halo scales.

  It would nevertheless be advantageous to have a framework that combines the strengths of both approaches: the physical interpretability and modular structure of the halo model, and the systematic description of non-linear clustering provided by HEFT. In this work, we present a hybrid EFT-halo model that addresses this challenge. Our approach constructs a set of operator fields inspired by HEFT, measured directly from simulations after collapsing particles belonging to each halo to its centre, thereby isolating the inter-halo clustering signal relevant for the 2-halo term, and eliminating any effect of the 1-halo term from the standard HEFT expansion.

  We will test the ability of this approach to produce accurate predictions for the two-point correlators of different tracers, by considering fields with different halo mass weighting schemes designed to mimic various astrophysical quantities, such as the thermal gas energy or thermal pressure (which governs the SZ effect); the star formation rate (governing the CIB); or the galaxy abundance through a halo occupation distribution (HOD). These probes provide a stringent test of the model, as they are sensitive to different halo mass ranges and clustering regimes.

  The method presented here complements other approaches presented in the literature to address the inaccuracy of the standard halo model in the 1-halo to 2-halo transition regime. These include the direct measurement and emulation of the halo-halo power spectrum \cite{1811.09504}, or its ratio with its linear bias prediction \cite{2011.08858}, as well as alternative halo model prescriptions, such as the Web-Halo Model of \cite{2508.10902v2}. The method is also reminiscent of other approaches presented in the literature \citep[e.g.][]{2004.09515,2104.10676}, in which the two-halo term is predicted through perturbative approaches. While different methods achieve varying levels of accuracy, the approach presented here offers a number of advantages. First, the bias parameters are not treated as free quantities, but are instead calibrated as a function of halo mass and cosmology using $N$-body simulations (e.g. exploiting the probabilistic bias framework of \citealt{2405.01950}, as we will do here). This allows us to construct a highly predictive model with virtually no free parameters describing the clustering of haloes. Secondly, the approach isolates the calculation of the 1-halo term, where all the non-linear modelling of complex astrophysical quantities naturally takes place.

  This paper is organised as follows. In Section~\ref{sec:th}, we present the theoretical framework, including the construction of the collapsed operator fields and their incorporation into the halo model. In Section~\ref{sec:matter}, we validate the model on the matter power spectrum. Section~\ref{sec:non_matter} presents the results for weighted observables and discusses the limitations of the baseline model. We discuss the interpretation and outlook in Section~\ref{sec:disc}.


\section{Theory}\label{sec:th}
  \subsection{The halo-model decomposition}\label{sec:th.concept}
    The halo model is based on the assumption that the non-linear density field can be described as a superposition of haloes \citep{astro-ph/0001493,astro-ph/0005010,astro-ph/0206508,2303.08752}. For the matter field, this may be written schematically as
    \begin{equation}
      \rho_m({\bf x}) = \sum_a \rho_m({\bf x}-{\bf x}_a|M_a),
    \end{equation}
    where $a$ labels haloes, $M_a$ is the halo mass, ${\bf x}_a$ is the halo centre, and $\rho_m({\bf x}|M)$ is the matter profile normalised such that $\int d^3x\,\rho_m({\bf x}|M)=M$. The same logic can be applied to any quantity $U$ associated with haloes. In that case, one writes
    \begin{equation}
      U({\bf x}) = \sum_a U(|{\bf x}-{\bf x}_a||M_a),
    \end{equation}
    where $U(r|M)$ is the profile of quantity $U$ around a halo of mass $M$. Often it may be convenient to write halo profiles as 
    \begin{equation}\label{eq:U_WU}
      U(r|M)=\bar{U}(M)\,W_U(r|M),
    \end{equation}
    where $\bar{U}(M)$ is the volume integral of the quantity $U$ associated with a halo of mass $M$, and $W_U$ is its spatial profile, normalised to unity when integrated over volume. Matter, galaxy number density, the thermal pressure traced by the Sunyaev--Zeldovich effect, the emissivity associated with the Cosmic Infrared Background (CIB), or X-ray emission can therefore all be described within the same formal structure, differing only in the mass dependence and spatial distribution of the relevant quantity.

    This construction separates the problem into three ingredients. The first ingredient is the halo mass function, $\bar n(M)$. The second is the Fourier-space profile of the quantity under consideration. For a spherically averaged profile, we define
    \begin{equation}\label{eq:prof_fourier}
      U(k|M) \equiv 4\pi\int dr\,r^2\frac{\sin kr}{kr}U(r|M).
    \end{equation}
    Note that, with our notation, the Fourier-space or real-space nature of a given function is determined by its argument ($k$ or $r$, respectively). Note also that the volume integral of $U$ can be recovered in the limit $\bar{U}(M)=U(k=0|M)$ (or, in other words, the normalised profiles satisfy $W_U(k=0|M)=1$). The third ingredient is the power spectrum of halo centres in two mass bins, $P_{hh}(k|M_1,M_2)$. The central goal of this work is to improve the modelling of this third ingredient, while retaining the standard halo-model treatment of mass integration and profiles.

    Throughout this paper we use $\bar n(M)$ to denote the differential halo abundance per logarithmic mass interval,
    \begin{equation}
      \bar n(M) \equiv \frac{d\bar n_h}{d\ln M}=
      M\frac{d\bar n_h}{dM}.
    \end{equation}
    Thus $\bar n(M)d\ln M$ is the number density of haloes in the
    interval $d\ln M$.
    
    For clustering statistics we work with fractional fluctuations of the corresponding fields. We therefore define the mean abundance, density, or intensity of the quantity $U$ as
    \begin{equation}
      \bar U=\int d\ln M\,\bar n(M)\,U(k=0|M),
    \end{equation}
    and introduce the normalised halo contribution
    \begin{equation}
      \tilde U(k|M)\equiv \frac{U(k|M)}{\bar U}.
    \end{equation}
    For the matter field, this reduces to $\tilde U_m(k|M)=M u_m(k|M)/\bar\rho_m$, with $u_m(k\to0|M)=1$.

  \subsection{The 1-halo and 2-halo terms}\label{ssec:th.hm_split}
    The power spectrum follows from pair counting. A pair of points contributing to the correlation of two quantities $U$ and $V$ can either belong to the same halo or to two different haloes. This gives the usual decomposition
    \begin{equation}
      P_{UV}(k)=P^{1\mathrm{h}}_{UV}(k)+P^{2\mathrm{h}}_{UV}(k),
      \label{halo_model_split}
    \end{equation}
    where $P^{1\mathrm{h}}_{UV}$ is the same-halo contribution and $P^{2\mathrm{h}}_{UV}$ is the different-halo contribution.

    The 1-halo term depends on the internal distribution of the relevant quantities within individual haloes. In a fully predictive halo model, this term requires a model for the profiles, occupation statistics, and their stochasticity. In this paper we do not introduce a new model for the 1-halo term. Instead, because our aim is to isolate and test the modelling of the 2-halo contribution, we measure $P^{1\mathrm{h}}_{UV}$ directly from the simulation using intra-halo pair counts, as described in Section~\ref{ssec:th.1h}. This allows us to focus on the transition regime without mixing errors in halo-centre clustering with errors in the internal halo model.

    The 2-halo term correlates two different halo centres, each dressed by the profile of the quantity being modelled. In the continuum halo-model notation it is
    \begin{align}
      P^{2h}_{UV}(k) ={}& \int d\ln M_1\,d\ln M_2\,\bar n(M_1)\bar n(M_2) \nonumber \\ & \times \tilde U(k|M_1)\tilde V(k|M_2)\,P_{hh}(k|M_1,M_2).
      \label{eq:hm_2h_general}
    \end{align}
    The factors $\tilde U(k|M)$ and $\tilde V(k|M)$ contain both the profile dependence and the normalisation by the mean value of the corresponding field. The appearance of these profile factors does not double count the 1-halo term. The pair in Eq.~\eqref{eq:hm_2h_general} is still a pair of two different haloes; the profile factors simply convolve the halo-centre field with the mean spatial distribution of $U$ and $V$ around each centre. If the observable is treated as concentrated at halo centres, one may set $\tilde U(k|M)=\bar U(M)/\bar U$. Retaining the full $k$ dependence accounts for the finite spatial extent of the observable around each halo.

    In the simulation implementation used below, the same expression can be written as a discrete sum over halo mass bins,
    \begin{equation}
    P^{2h}_{UV}(k) = \sum_{m,n}\frac{N_mN_n}{V_{\rm box}^2}\, \tilde U_m(k)\tilde V_n(k)\, P^{mn}_{hh}(k),
    \label{eq:discrete_2h}
    \end{equation}
    where $N_m$ is the number of haloes in the $m$-th mass bin, $\tilde U_m(k)$ is the mean contribution of haloes in that bin to the fractional $U$ field, and $V_{\rm box}$ is the simulation volume.

  \subsection{The standard 2-halo approximation and its limitations}\label{ssec:th.standard_2h}
    Up to this point, the halo model is a formal decomposition. It becomes predictive once one specifies the halo mass function, halo profiles, and halo-centre clustering. The standard closure for the latter is to approximate the halo-halo power spectrum by a linear-bias rescaling of the linear matter power spectrum,
    \begin{equation}
      P_{hh}(k|M_1,M_2)\simeq
      b^{\rm E}_1(M_1)b^{\rm E}_1(M_2)P_{\rm lin}(k),
      \label{eq:linear_halo_bias}
    \end{equation}
     where $b^{\rm E}_1(M)$ is the Eulerian linear halo bias. This approximation has two distinct consequences. The first is the familiar one: it replaces the non-linear clustering of halo centres by the linear matter power spectrum. The second is more structural. At fixed $k$, Eq.~\eqref{eq:linear_halo_bias} predicts that the halo-halo power spectrum as a function of halo mass is an outer product. Equivalently, after subtracting shot noise, the cross-correlation coefficient between two halo samples,
    \begin{equation}
    r_{12}(k)\equiv
    \frac{P_{hh}(k|M_1,M_2)}
    {\left[P_{hh}(k|M_1,M_1)P_{hh}(k|M_2,M_2)\right]^{1/2}},
    \label{eq:halo_cross_correlation_coefficient}
    \end{equation}
    is forced to satisfy
    \begin{equation}
    r_{12}^{\rm lin}(k)=1,
    \label{eq:linear_halo_r_one}
    \end{equation}
    for all halo masses with positive Eulerian bias. This conclusion would not be changed by replacing $\smash{b^{\rm E}_1(M)}$ with a deterministic scale-dependent bias $b^{\rm E}(k,M)$: the model would still be proportional to the product of one amplitude for $M_1$ and one amplitude for $M_2$. Thus the standard 2-halo approximation assumes that all halo samples trace the same underlying large-scale field with perfect coherence, differing only in their bias amplitudes.

    Substituting Eq.~\eqref{eq:linear_halo_bias} into Eq.~\eqref{eq:hm_2h_general} gives the familiar factorised 2-halo term,
    \begin{align}
    P^{2h,{\rm lin}}_{UV}(k) ={}& \left[ \int d\ln M\,\bar n(M)b^E_1(M)\tilde U(k|M) \right] \nonumber\\& \times
    \left[\int d\ln M\,\bar n(M)b^E_1(M)\tilde V(k|M)\right] P_{\rm lin}(k). \label{eq:standard_2h_factorised} \end{align}
    This expression captures the asymptotic large-scale behaviour. However, it is restrictive in two ways. First, it uses the linear matter power spectrum to describe halo-centre clustering in a regime where non-linear evolution and non-linear bias are already important. Second, and more fundamentally, it makes the halo-halo power-spectrum matrix in mass space fully factorised. It therefore cannot describe imperfect coherence between different halo populations, even if the bias amplitude is allowed to become scale dependent. These restrictions are among the main sources of inaccuracy of the standard halo model in the transition between the 1-halo and 2-halo regimes, where non-linear bias, halo exclusion, and stochasticity all become relevant \citep{astro-ph/0411777,1406.5013,2011.08858}.

    The standard halo model also obeys important large-scale consistency conditions. For the matter field, mass conservation requires
    \begin{equation}
      \int d\ln M\,\bar n(M)\frac{M}{\bar\rho_m}=1,
      \label{eq:mass_conservation_hm}
    \end{equation}
    and the large-scale matter field must be unbiased,
    \begin{equation}
      \int d\ln M\,\bar n(M)\frac{M}{\bar\rho_m}b^{\rm E}_1(M)=1.
      \label{eq:eulerian_bias_consistency}
    \end{equation}
    Equivalently, since $b^{\rm E}_1=1+b^{\mathcal{L}}_1$, the Lagrangian linear-bias contribution must satisfy
    \begin{equation}
      \int d\ln M\,\bar n(M)\frac{M}{\bar\rho_m}b^{\mathcal{L}}_1(M)=0.
      \label{eq:lagrangian_bias_consistency}
    \end{equation}
    These constraints will become useful in Section~\ref{sec:matter}, where we test whether the CHEFT construction recovers the matter power spectrum when the measured 1-halo term is combined with our modified 2-halo term.

  \subsection{Replacing the standard 2-halo term with CHEFT}\label{ssec:th.hm}
    The purpose of the CHEFT construction is to replace Eq.~\eqref{eq:linear_halo_bias}, while leaving the rest of the halo-model structure intact. We retain the decomposition of Eq.~\eqref{halo_model_split}, the mass integration in Eq.~\eqref{eq:hm_2h_general}, and the profile factors $U(k|M)$ and $V(k|M)$. We also retain a separate 1-halo contribution, which in this paper is measured directly from the simulation. The only ingredient that is changed is the model for the halo-centre power spectrum $P_{hh}(k|M_1,M_2)$.

    Our replacement is inspired by Hybrid Effective Field Theory (HEFT; \citealt{1910.07097,2101.11014,2103.09820,2109.08699,2107.10287,2101.12187}). In HEFT, tracer fields are written as a Lagrangian-bias expansion in operators defined in the initial conditions, and these weighted fields are advected to Eulerian space using the fully non-linear displacement field measured from simulations. This captures a large fraction of the non-linear evolution while preserving the interpretability of a bias expansion. However, the standard HEFT fields contain correlations between particles within the same halo. They therefore cannot be inserted directly into the halo model without double counting the 1-halo contribution.

    We solve this by constructing a collapsed HEFT basis, or CHEFT basis. Particles are weighted by Lagrangian operators as in HEFT, but particles belonging to the same halo are collapsed to the halo centre (defined as the position of the most-bound particle) before the operator spectra are measured. This removes intra-halo structure from the HEFT templates and makes them suitable for modelling the 2-halo term. The corresponding closure for halo-centre clustering is
    \begin{equation}
      P_{hh}(k|M_1,M_2)=
      \sum_{i,j}b^{\mathcal{L}}_i(M_1)b^{\mathcal{L}}_j(M_2)P^{ij}_{\rm CHEFT}(k),
      \label{eq:cheft_expansion}
    \end{equation}
    with $b^{\mathcal{L}}_{O_1}(M)=1$ for the constant operator. This equation is the direct analogue of Eq.~\eqref{eq:linear_halo_bias}. The standard model uses one bias coefficient and one template, $P_{\rm lin}(k)$; CHEFT uses several Lagrangian response coefficients and a matrix of non-linear collapsed-operator templates.

    In matrix language, Eq.~\eqref{eq:cheft_expansion} replaces the fully factorised mass dependence of the standard 2-halo term by a finite-dimensional operator expansion. At fixed $k$, the prediction may be written schematically as
    \begin{equation}
      P_{hh}(k|M_1,M_2)={\bf b}^{\mathcal{L}}(M_1)^{\rm T}{\bf P}_{\rm CHEFT}(k){\bf b}^{\mathcal{L}}(M_2),
      \label{eq:cheft_matrix_form}
    \end{equation}
    where ${\bf b}^{\mathcal{L}}(M)$ is the vector of Lagrangian bias coefficients and ${\bf P}_{\rm CHEFT}(k)$ is the matrix of collapsed-operator spectra. The cross-correlation coefficient between two halo masses is therefore not forced to be unity. This is the sense in which CHEFT generalises the standard 2-halo term: it allows different halo populations to project differently onto several non-linear fields, rather than assuming that all masses trace a single common field with different amplitudes.

    It is important to distinguish the two factors entering Eq.~\eqref{eq:cheft_expansion}. The spectra $\smash{P^{ij}_{\rm CHEFT}(k)}$ are empirical power spectra of collapsed operator fields, measured from the simulation. 
    These quantities play the role of a non-linear basis for halo-centre clustering:
    the mass dependence of halo clustering enters through the coefficients $b^{\mathcal{L}}_i(M)$, which specify how haloes of mass $M$ respond to the different Lagrangian operators. The improvement over the standard halo model is therefore not only that the templates are non-linear, but also that the mass dependence of $P_{hh}$ is no longer constrained to be a single outer product.
    It may seem that the expansion in Eq.~\eqref{eq:cheft_expansion} is unnecessary, as the required $P_{hh}(k|M_1,M_2)$ can be measured directly from the simulation. However, in practice this would need to involve coarse bins in mass, whereas the operator-based expansion gives results that vary smoothly with mass. But this approach creates a problem by introducing the unknown nuisance parameters $b^{\mathcal{L}}_i(M)$. These could be left to float in any data analysis, but this would reduce the precision of any cosmological inference. We therefore explain below in Section~\ref{sssec:th.hm.bias} how it is possible to obtain unique optimal choices for these parameters.

    Substituting Eq.~\eqref{eq:cheft_expansion} into Eq.~\eqref{eq:hm_2h_general} gives
    \begin{equation}
      P^{2\mathrm{h}}_{UV}(k)=
      \sum_{i,j}I^i_U(k)I^j_V(k)P^{ij}_{\rm CHEFT}(k),
      \label{eq:pk_2h_cheft}
    \end{equation}
    where
    \begin{align}
      I^i_U(k) & \equiv \int d\ln M\,\bar n(M)b^{\mathcal{L}}_i(M)\tilde U(k|M),
      \label{eq:IU_HM}\\
      & = \sum_n \frac{N_n}{V_{\rm box}}\, b^{\mathcal{L}}_{i,n}\, \tilde U_n(k).
      \label{eq:weighted_integral}
    \end{align}
    Equation~\eqref{eq:pk_2h_cheft} is the working expression used in this paper. It preserves the halo-model separation between profiles, mass weighting, and halo-centre clustering, but replaces the linear 2-halo approximation by a simulation-calibrated non-linear operator expansion.

  \begin{figure*}
      \includegraphics[width=\textwidth]{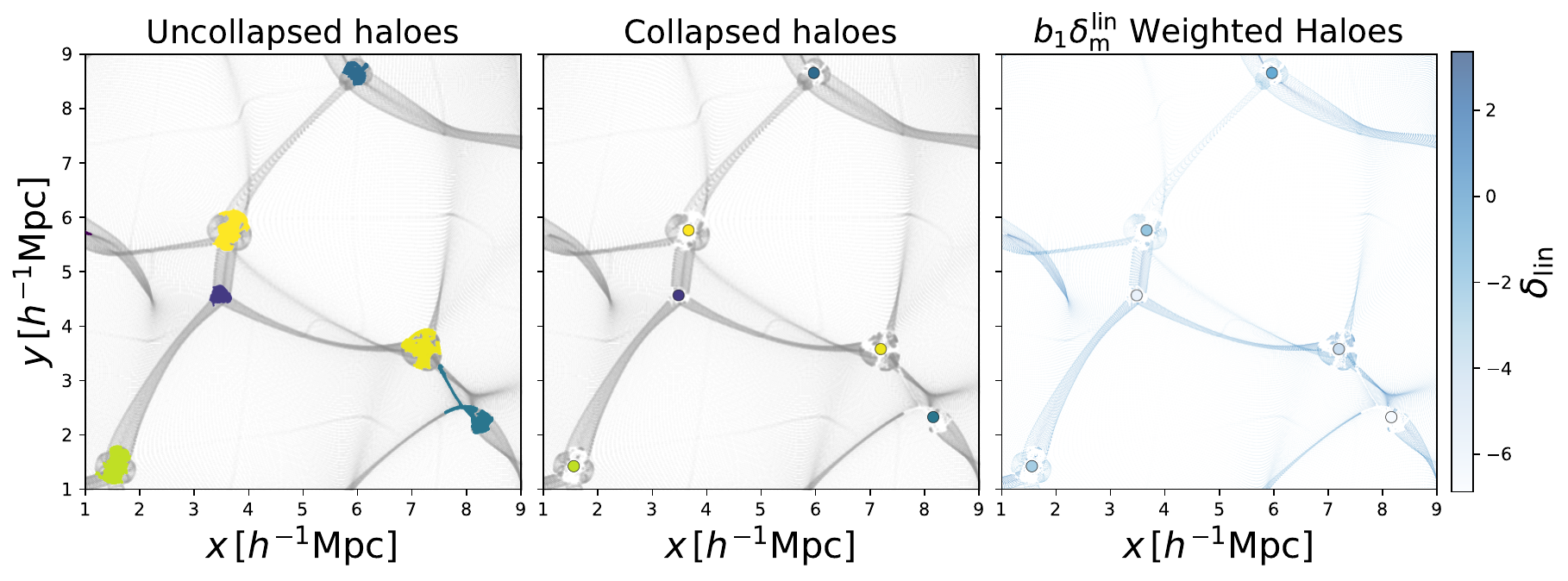}
      \caption{Schematic illustration of the construction of the collapsed operator fields used in this work. In the standard particle field, correlations include both intra-halo and inter-halo contributions. In the CHEFT construction, all particles belonging to a given halo are displaced to the halo centre, while field particles are left at their Eulerian positions and treated as unresolved objects. The resulting collapsed fields preserve the large-scale Lagrangian operator weighting but remove the internal halo structure. The CHEFT spectra measured from these fields are therefore suitable as templates for the 2-halo part of the halo model.}
      \label{fig:concept}
    \end{figure*}

  \subsection{Collapsed operator fields}\label{ssec:th.cheft}
    We now describe the construction of the CHEFT templates entering Eq.~\eqref{eq:cheft_expansion}. The construction starts from the same Lagrangian operators used in the second-order HEFT bias expansion. We define
    \begin{align}
      O_1({\bf q}) &\equiv 1, &
      O_\delta({\bf q}) &\equiv \delta_{\rm L}({\bf q}),\nonumber\\
      O_{\delta^2}({\bf q}) &\equiv \delta^2_{\rm L}({\bf q})-\langle\delta^2_{\rm L}\rangle, &
      O_{s^2}({\bf q}) &\equiv s^2({\bf q})-\langle s^2\rangle,\nonumber\\
      O_{\nabla^2\delta}({\bf q}) &\equiv \nabla^2\delta_{\rm L}({\bf q}),
      \label{eq:lagrangian_operators}
    \end{align}
    where $\delta_{\rm L}$ is the linear density field, $s^2\equiv s_{ij}s^{ij}$, and the trace-free tidal tensor is defined in Fourier space as
    \begin{equation}
      s_{ij}({\bf k})\equiv
      \left(\frac{k_i k_j}{k^2}-\frac{1}{3}\delta^{\rm K}_{ij}\right)\delta_{\rm L}({\bf k}).
      \label{eq:tidal_tensor}
    \end{equation}
    The mean subtractions define the normal-ordered operators, ensuring that the second-order fields have zero mean \citep{1805.05304}.

    Each simulation particle carries the values of these operators evaluated at its initial Lagrangian position ${\bf q}_p$. The CHEFT step is to map these Lagrangian weights to Eulerian space after removing intra-halo structure. For each resolved halo, all particles belonging to that halo are assigned the Eulerian position of the halo centre. Particles not belonging to any resolved halo are left at their Eulerian positions and are treated as unresolved objects of mass equal to the particle mass. Thus, for a collapsed object $g$ located at ${\bf x}_g$, we define the operator weight
    \begin{equation}
      W_i^g \equiv \sum_{p\in g}O_i({\bf q}_p),
      \label{eq:collapsed_weight}
    \end{equation}
    where the sum runs over all particles in the halo, or over a single particle for the unresolved field-particle objects. The corresponding collapsed operator field is
    \begin{equation}
      \delta^{\rm CHEFT}_i({\bf x}) =
      \frac{1}{\bar n_p}\sum_g W_i^g\delta_{\rm D}({\bf x}-{\bf x}_g)
      -\langle O_i\rangle,
      \label{eq:collapsed_operator_field}
    \end{equation}
    where $\bar n_p$ is the mean number density of the particle catalogue used to construct the fields.

    This definition uses the sum of the Lagrangian operator values over the proto-halo patch, rather than evaluating the operator at a single Lagrangian centre-of-mass position. This choice is deliberate. The bias expansion describes the response of the abundance of a finite Lagrangian patch to its large-scale environment; the particle sum is therefore the natural discretised version of the proto-halo average. Evaluating the operator at a single Lagrangian point would discard information about the finite extent of the proto-halo and would not correspond to the same mass-weighted field that enters the halo-model decomposition.

    An important consequence of this construction is that the CHEFT basis is tied to the halo catalogue used to define the collapse. In particular, changing the mass resolution changes the boundary between resolved collapsed objects and unresolved field particles: particles that are left uncollapsed in a lower-resolution catalogue may become members of newly resolved haloes at higher resolution. This does not imply that the model is ill-defined, but it means that the CHEFT basis, like the standard halo-model ingredients, should be understood at fixed halo definition and mass resolution. The expected impact of newly resolved low-mass haloes is controlled by their physical size. Replacing the internal particle distribution of a halo of characteristic radius $R$ by its centre changes its Fourier-space contribution only through terms that are suppressed on scales $kR\ll 1$, i.e. through profile-like or higher-derivative corrections. Thus, for a catalogue complete above a minimum resolved halo radius $R_{\rm min}$, the effect of changing the treatment of smaller unresolved objects is expected to be small on scales satisfying $kR_{\rm min}\ll 1$. We do not attempt a resolution-convergence study in this proof-of-concept analysis; such a test will be required for an emulator-level implementation, where the CHEFT spectra, bias functions, mass function, and 1-halo prescription must all be defined consistently for a specified halo catalogue (see e.g. \citealt{2510.13962}).

    The CHEFT basis spectra are then defined by
    \begin{equation}
      \langle\delta^{\rm CHEFT}_i({\bf k})\delta^{\rm CHEFT}_j(-{\bf k}')\rangle
      =(2\pi)^3\delta_{\rm D}({\bf k}-{\bf k}')P^{ij}_{\rm CHEFT}(k).
      \label{eq:cheft_power_definition}
    \end{equation}
    Since all CHEFT fields are built from the same collapsed catalogue, there is is a non-zero shot-noise contribution for both cross- and auto-spectra. In practice, we subtract this contribution by measuring the power spectra of randomised catalogues with the same positions randomised and the same operator weights, and averaging over random realisations. Equivalently, for a Poisson catalogue with fixed weights, the expected contribution is
    \begin{equation}
      P^{ij}_{\rm SN}=\frac{1}{\bar n_p^2\mathcal V}\sum_g W_i^gW_j^g.
      \label{eq:cheft_shot_noise}
    \end{equation}
    After this subtraction, the spectra $P^{ij}_{\rm CHEFT}$ contain correlations between distinct collapsed objects and provide the 2-halo templates used in Eq.~\eqref{eq:cheft_expansion}.

    The construction is illustrated in Fig.~\ref{fig:concept}. For the five operators in Eq.~\eqref{eq:lagrangian_operators}, there are 15 distinct auto- and cross-spectra. These are the CHEFT basis spectra used throughout this work. Appendix~\ref{Appendix:basis_terms} compares these collapsed spectra with the corresponding uncollapsed HEFT spectra, showing explicitly how the collapse removes intra-halo contributions.

   \begin{figure}
      \includegraphics[width=0.98\columnwidth]{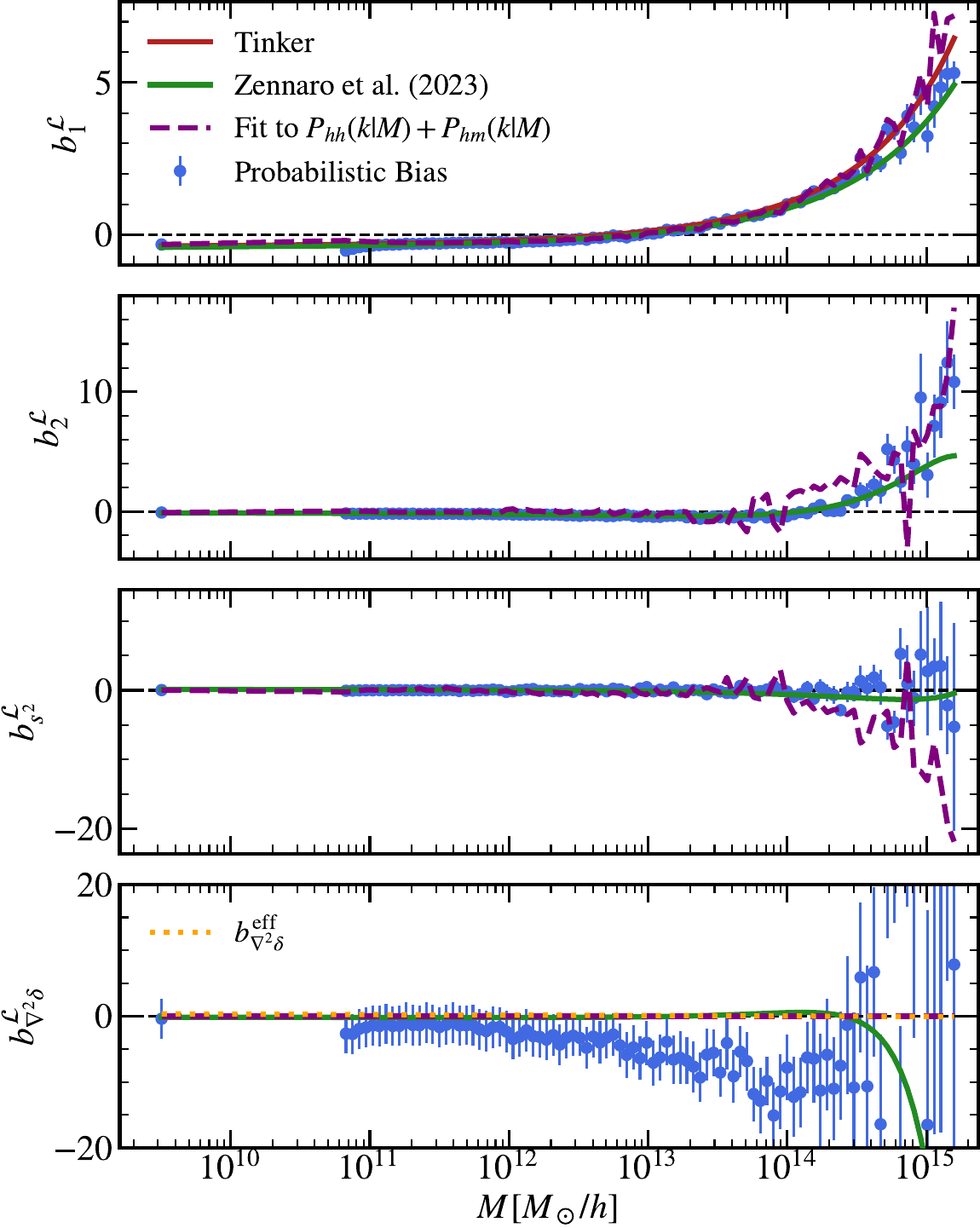}
      \caption{Halo bias parameters as a function of halo mass measured from the probabilistic bias approach. Circles show the bias functions $b_i(M)$ obtained from the probabilistic method, while solid lines correspond to the fitting functions of Tinker et al., and to the relations presented in Zennaro et al. (2023). The dashed lines indicate the best-fitting bias parameters obtained from direct fits to the auto-spectra $P_{hh}(k|M,M)$ described in Appendix~\ref{Appendix:Phh_fit}. The probabilistic bias measurements yield well-behaved bias relations across the full mass range, in good agreement with previous results. The comparison with the direct fits to $P_{hh}(k|M,M)$ shows that the probabilistic bias relations provide a consistent description of halo clustering at the level of individual mass bins, supporting their use as inputs to the CHEFT-based halo model.}
      \label{fig:ProbBias}
    \end{figure}
    
  \subsection{Halo bias functions from probabilistic bias}\label{sssec:th.hm.bias}
    The second ingredient required by Eq.~\eqref{eq:cheft_expansion} is the set of mass-dependent Lagrangian bias functions $b^{\mathcal{L}}_i(M)$. In the standard halo model, the mass function and the linear halo bias are usually calibrated from simulations. Here we follow the same philosophy, but extend it to the full set of Lagrangian response coefficients entering the CHEFT expansion. To keep the baseline model predictive, we do not fit these coefficients to the power spectra analysed in this paper. Instead, we use independent measurements obtained with the probabilistic-bias framework of \citet{2405.01951,2405.01950}.

    This framework is based on the peak-background-split interpretation of bias, in which halo abundance responds to long-wavelength perturbations of the initial density field \citep{Kaiser1984,Bardeen1986,0202393,1611.09787}. If only a constant large-scale density perturbation $\delta_0$ is considered, the response of the halo abundance can be written as
    \begin{equation}
      F(\delta_0)\equiv\frac{n_h(\delta_0)}{n_{h,0}}
      =1+b^{\mathcal{L}}_1\delta_0+\frac{1}{2}b^{\mathcal{L}}_2\delta_0^2+\cdots,
      \label{eq:pbs_response}
    \end{equation}
    so that
    \begin{equation}
      b^{\mathcal{L}}_n=
      \left.\frac{\partial^nF(\delta_0)}{\partial\delta_0^n}\right|_{\delta_0=0}.
      \label{eq:pbs_bias_definition}
    \end{equation}

    The probabilistic-bias method estimates these response coefficients from the distribution of Lagrangian environments sampled by haloes. In the density-only case, let $p(\delta)$ be the probability distribution of the smoothed linear density field, and let $p(\delta|h)$ be the same distribution evaluated at the Lagrangian positions of haloes in a given mass bin. The ratio
    \begin{equation}
      f(\delta)\equiv\frac{p(h|\delta)}{p(h)}=\frac{p(\delta|h)}{p(\delta)}
      \label{eq:pb_ratio}
    \end{equation}
    measures the excess probability of finding a halo in an environment with linear density $\delta$ \citep{2405.01951,2405.01950}. A long-wavelength perturbation shifts the local environment, giving
    \begin{equation}
      F(\delta_0)=\int d\delta\,p(\delta)f(\delta+\delta_0).
      \label{eq:pb_shift}
    \end{equation}
    Taking derivatives with respect to $\delta_0$ and integrating by parts yields
    \begin{equation}
      b^{\mathcal{L}}_n=(-1)^n\int d\delta\,\frac{p^{(n)}(\delta)}{p(\delta)}p(\delta|h),
      \label{eq:pb_density_bias}
    \end{equation}
    where $p^{(n)}$ denotes the $n$-th derivative of the unconditional density distribution.

    The extension to the full bias basis is multivariate \citep{1611.09787,2012.09837,2405.01950}. The density, tidal field, and higher-derivative operators are correlated, so they cannot be treated as independent one-dimensional variables. Instead, the relevant object is the joint distribution of the smoothed Lagrangian environment,
    \begin{equation}
      {\bf y}=\{\delta,\nabla^2\delta,s_{ij},\ldots\},
      \label{eq:pb_environment}
    \end{equation}
    and its conditional counterpart at halo Lagrangian positions, $p({\bf y}|h)$. Equivalently,
    \begin{equation}
      f({\bf y})=\frac{p({\bf y}|h)}{p({\bf y})}.
      \label{eq:pb_multivariate_ratio}
    \end{equation}
    The scalar coefficients used in the CHEFT expansion, such as $\smash{b^{\mathcal{L}}_1}$, $b^{\mathcal{L}}_2$, $b^{\mathcal{L}}_{s^2}$, and $b^{\mathcal{L}}_{\nabla^2\delta}$, are obtained as derivatives or tensor projections of this multivariate response around a vanishing background perturbation. This is the step that allows tidal and higher-derivative bias parameters to be measured consistently, rather than fitted independently from clustering statistics.

    The tensorial probabilistic-bias measurements provide mass-dependent Lagrangian response coefficients for the same operator content used in our fiducial CHEFT expansion: the linear density, the quadratic density operator, the tidal invariant, and the Laplacian. We therefore identify the measured coefficients with $\left\{ b^{\mathcal{L}}_1(M),\,b^{\mathcal{L}}_2(M),\,b^{\mathcal{L}}_{s^2}(M),\, b^{\mathcal{L}}_{\nabla^2\delta}(M)\right\}$.
    In the notation of \citet{2405.01950}, these coefficients are associated with the response variables $\{J_2,J_{22},J_{2=2},J_4\}$, respectively. The probabilistic-bias estimator also includes spatial-correction terms designed to reduce the residual dependence on the smoothing scale. These terms enter only as part of the bias-estimation procedure; they are not included as additional independent operators in the CHEFT basis. The explicit definitions of the estimators are given in \citet{2405.01950}.

    The probabilistic-bias construction requires the linear density field to be smoothed on scales where the peak-background-split description is meaningful \citep{Kaiser1984,Peacock&Heavens1985,Bardeen1986,Coles1986,Cole&Kaiser1989,1611.09787,2405.01950}. We use a sharp-$k$ filter and adopt a smoothing scale $k^{\rm PB}_s=0.3\,h\,{\rm Mpc}^{-1}$ for the fiducial measurements. The resulting coefficients are measured independently in halo mass bins and then used as empirical functions of mass, $b^{\mathcal{L}}_i(M)$.

    The measurements are shown in Fig.~\ref{fig:ProbBias}, together with previous fitting functions and the bias values obtained from direct fits to the halo auto-spectra discussed in Appendix~\ref{Appendix:Phh_fit}. The agreement is generally good over the mass range relevant for this work, with the largest deviations appearing at the high-mass end, where the number of haloes is small, and at the low-mass end, where resolution and halo-finder effects become more important. The uncertainties are estimated using jackknife resampling over 64 sub-volumes of the simulation box.

    Among the measured bias functions, $b^{\mathcal{L}}_{\nabla^2\delta}(M)$ requires special care. The Laplacian operator is a higher-derivative contribution and is especially sensitive to the coarse-graining scale used to define the linear density field. In the baseline model we therefore exclude it and define
    \begin{equation}
      \mathrm{CHEFTmin}=\{1,\delta_{\rm L},\delta_{\rm L}^2,s^2\}.
      \label{eq:cheftmin_definition}
    \end{equation}
    This is the default predictive model used below. We later introduce an extended model, CHEFText, in which the Laplacian contribution is restored as an effective higher-derivative term calibrated from simulations. Section~\ref{sec:matter} motivates this choice using the matter consistency relations, while Section~\ref{ssec:non_mater.extended} quantifies the improvement obtained with the effective Laplacian model.

  \subsection{Simulated tracers and halo profiles}\label{ssec:th.wM}
    One of the main strengths of the halo model is that, once the halo-centre clustering is specified, different observables can be generated by changing the mass dependence and profile of the relevant quantity. In this work, the scale-dependent part of the halo profile ($W_U(k|M)$ in Eq. \ref{eq:U_WU}) is measured directly from the simulation by spherically averaging and binning each halo. Since the 1-halo term is also measured directly, the purpose of these profiles here is to test the 2-halo part of the model in Eq.~\eqref{eq:pk_2h_cheft}, not to provide a complete analytic description of the internal astrophysics of each tracer.

    In a fully predictive implementation, the measured 1-halo term used in this work would be replaced by the standard halo-model expression
    \begin{equation}
      P^{1h}_{UV}(k) = \int d\ln M\,\bar n(M)\, \tilde U(k|M)\tilde V(k|M),
      \label{eq:1halo_analytic}
    \end{equation}
    for deterministic halo profiles and weights. More generally, this expression must be extended to include occupation moments, profile-to-profile scatter, and the covariance between the quantities $U$ and $V$ at fixed halo mass. These ingredients are observable dependent and are not the focus of this proof-of-concept study. We therefore measure $P^{1h}_{UV}$ directly from the simulation, as explained in the next subsection, so that the validation isolates the accuracy of the CHEFT prediction for the 2-halo contribution.

    The decomposition in Eq.~\eqref{eq:U_WU} is the simplest halo-model description, in which the relevant observable is specified by the dependence of its amplitude on halo mass and by an average profile at fixed mass. This is sufficient for the 2-halo calculation if the residual profile-to-profile variations at fixed mass are uncorrelated with the large-scale fields that determine halo clustering. In general, however, this is an additional simplifying assumption. Halo profiles and observable weights may depend on secondary properties such as concentration, formation time, spin, orientation, or tidal environment, and these quantities may themselves correlate with the large-scale density and tidal fields. A fully predictive model would then require extending the halo-model integrals from mass alone to a larger set of halo properties, or equivalently including assembly-bias and profile-alignment terms. In the proof-of-concept tests presented here we deliberately restrict the weights to functions of halo mass alone, so that the accuracy of the CHEFT prediction for the mass-dependent 2-halo term can be isolated. Note that technically our framework allows for the incorporation of assembly bias at the 2-halo level. If the bias coefficients were permitted to differ from the fiducial values, the bias expansion allows for an adjustable influence of beyond halo-mass terms on halo clustering. However, we are interested in a fully predictive approach, with as few free parameters as possible, so that the model has optimal cosmological parameter information.

    To construct mock observations of non-matter tracers, we apply halo-mass-dependent weights to simulation particles. For a quantity with total halo weight $\bar U(M)$, the particle weight is
    \begin{equation}
      w^U_p\equiv \bar U(M_p)\frac{m_p}{M_p},
      \label{eq:particle_weight}
    \end{equation}
    where $m_p$ is the particle mass and $M_p$ is the mass of the halo containing the particle. This normalisation ensures that the total weight assigned to a halo of mass $M_h$ is $\sum_{p\in h}w^U_p=\bar U(M_h)$. Particles not associated with any resolved halo are treated as objects of mass $m_p$, consistently with the treatment adopted in the 2-halo term. Because we do not perturb the particle positions within haloes, the profile shape for these mock observables is that of the simulated matter distribution. This does not reproduce the true small-scale profiles of SZ, CIB, or galaxy observables, but it is sufficient for validating the CHEFT prediction for the 2-halo contribution.
    
    We consider three representative weighting schemes:
    
    \begin{itemize}
        \item \textbf{Sunyaev--Zeldovich (SZ)-like weights.} The thermal SZ effect traces the thermal gas pressure $P_{\rm th}$. The associated normalisation factor $\bar{U}(M)$ is then proportional to the total thermal energy of the halo $E_{\rm th}\propto M\,T$, where $T$ is the gas temperature. If the gas is virialised, $k_{\rm B}T\sim GM/R\propto M^{2/3}$, where $k_{\rm B}$ is the Boltzmann constant, and $R$ is the virial radius of the halo. The SZ normalisation factor is therefore
        \begin{equation}
          \bar{U}_{\mathrm{SZ}}(M) \propto M^{1+\alpha},
        \end{equation}
        with $\alpha=2/3$, and an arbitrary normalisation factor.
        We will take advantage of this simple weighting scheme to test the accuracy of CHEFT not just for SZ, but over a continuous range of mass weighting schemes. We consider values of $\alpha$ in the range $[0,\,1]$, sampling both nearly uniform weighting ($\alpha \simeq 0$, corresponding to the matter field) and strongly mass-weighted regimes, dominated by the largest haloes ($\alpha \simeq 1$).
        \item \textbf{Cosmic Infrared Background (CIB)-like weights.} The CIB is dominated by the infrared emission heated by ultraviolet light from massive young stars and, as such, it tracks star formation \citep{1967ApJ...147..868P,astro-ph/0009151}. The normalisation factor for the CIB is therefore proportional to the total star formation rate (SFR) of the halo. We model this as a broken power law of the form
        \begin{equation}
          \bar{U}_{\mathrm{CIB}}(M) = A \frac{M/M_1}{(M_1/M)^{\beta} + (M/M_1)^{\gamma}},
        \end{equation}
        where we adopt $A=5$, $M_1 = 8.4 \times 10^{11}\,h^{-1}M_\odot$, $\beta = 3.4$, and $\gamma = 0.8$. This choice of parameters reproduces the best-fit model found by \citet{2209.05472}.
        \item \textbf{Galaxies and HOD-like weights.} Galaxy clustering tracks the relative fluctuations in the abundance of galaxies. As such, the normalisation factor tracks the total number of galaxies in a given halo. To model this here we use the HOD parametrisation of \citet{astro-ph/0408564,abs/1005.2413}, which considers the contributions from both central and satellite galaxies:
        \begin{equation}
          \bar{U}_{\mathrm{HOD}}(M)=N_c(M)+N_s(M),
        \end{equation}
        where
        \begin{align}
          &N_c(M)=\frac{1}{2}\left[1+\mathrm{erf}\left(\frac{\log_{10}M-\log_{10}M_{\min}}{\sigma_{\log M}}\right)\right],\\
          &N_s(M)=N_c(M)\left(\frac{M}{M_1}\right)^\alpha.
        \end{align}
        We adopt $M_{\min} = 10^{11.7}\,h^{-1}M_\odot$, $\sigma_{\log M} = 0.2$, $M_1 = 10^{13}\,h^{-1}M_\odot$, and $\alpha = 1$. These values are chosen as representative plausible values of realistic Hydrodynamical simulations \citep{2511.01441}. These parameters correspond to an Eulerian bias of $b_1^{\cal E}\approx 1.1$, which would be roughly typical of an Emission Line Galaxy clustering sample \citep{1911.05714}. We tested different choices for these parameters, finding qualitatively similar results in all cases.
      \end{itemize}

       \begin{figure}
      \includegraphics[width=0.98\columnwidth]{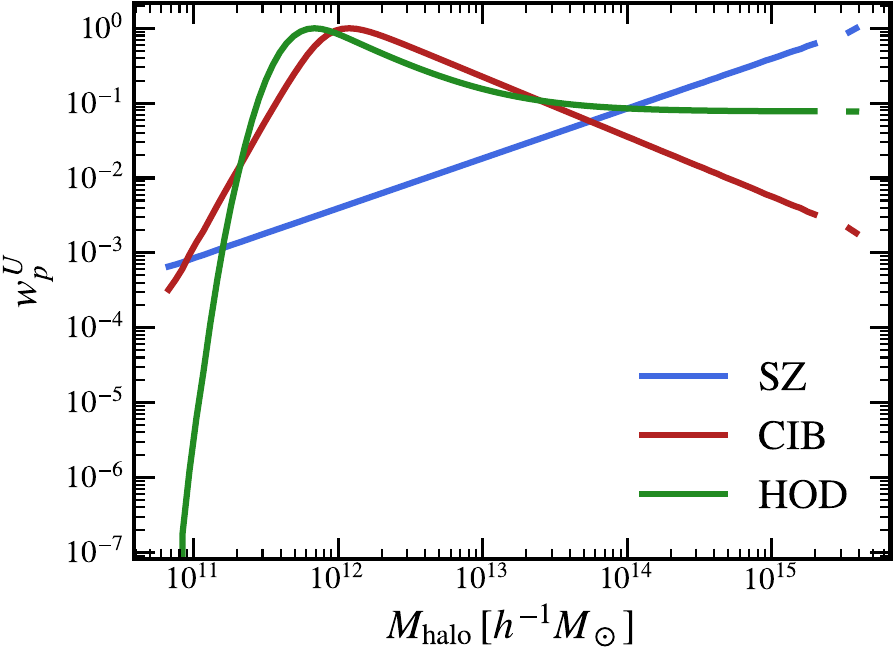}
      \caption{Per-particle weighting functions $w^U_p=\bar U(M)m_p/M$ used to construct the mock observables, shown as a function of halo mass and normalised to their maximum value for visual comparison. These are not the total halo weights $\bar U(M)$, but the weights assigned to each particle inside a halo of mass $M$. Thus, for example, an HOD with $\bar U_{\rm HOD}(M)\propto M$ at high mass corresponds to an approximately constant per-particle weight. The SZ-like weighting strongly emphasises high-mass haloes, while the HOD and CIB prescriptions probe a broader range of halo masses. This difference in mass sensitivity is a key factor in the performance of the model, as observables receiving comparable contributions from many halo masses provide a more stringent test of the halo-halo power spectrum.}
      \label{fig:weights}
    \end{figure}

    The particle weights are shown in Fig.~\ref{fig:weights}. Once the weighted mock fields are constructed, we measure their auto-spectra $P_{UU}(k)$ and cross-spectra with matter, $P_{Um}(k)$. These measurements are the target observables against which we compare the halo-model predictions built from the measured 1-halo term and the CHEFT 2-halo term.

  \subsection{Direct measurement of the 1-halo term}\label{ssec:th.1h}
  
    In all predictions shown in this paper, the 1-halo contribution is not modelled analytically. It is measured directly from the simulation, so that the validation focuses on the CHEFT description of the 2-halo term. We compute weighted intra-halo particle pair counts as a function of separation, $DD^{1h}_{UV}(r)$, using \textsc{Corrfunc}\footnote{\url{https://pypi.org/project/Corrfunc/}}\citep{1911.03545}. The sum includes only pairs of particles belonging to the same halo, with the appropriate weights for the two fields $U$ and $V$. These pair counts are normalised by the expected number of pairs in a uniform random distribution,
    \begin{equation}
      \xi^{1h}_{UV}(r)  =  \frac{DD^{1h}_{UV}(r)}{RR_{UV}(r)} ,
      \label{eq:xi_1h}
    \end{equation}
    where $RR_{UV}$ includes the same mean-density normalisation as the corresponding fractional fields. The 1-halo power spectrum is then obtained through the Hankel transform:
    \begin{equation}
      P^{1h}_{UV}(k)  =  4\pi \int dr\,r^2\, \xi^{1h}_{UV}(r)\,j_0(kr),
      \label{eq:p1h_hankel}
    \end{equation}
    where $j_0(x)=\sin x/x$.
    
    For the non-matter tracers considered in this work, we use the same pair-counting structure but apply the mass-dependent particle weights defined in Section~\ref{ssec:th.wM}. This allows us to evaluate the 1-halo term for all mock observables considered here without recomputing the pair counts for each weighting scheme.
    
    The measured 1-halo term should not be interpreted as a complete large-scale model on its own. A standard analytic 1-halo term behaves as a profile-filtered white-noise contribution on sufficiently large scales. If extrapolated to arbitrarily small $k$, this would eventually violate the mass- and momentum-conservation behaviour of the matter power spectrum, for which the stochastic contribution must be suppressed on large scales. This well-known limitation of the standard halo model is not relevant for the tests presented here, because over the range of scales shown the total prediction is dominated on large scales by the 2-halo term and the measured 1-halo contribution is used only to isolate the transition regime. A fully predictive implementation extending to much larger scales would require a compensated 1-halo prescription, or an equivalent treatment of halo exclusion and stochastic counterterms, consistent
    with the large-scale conservation constraints.
    
    In a future predictive application of CHEFT, the measured 1-halo term would therefore have to be replaced by a model for the relevant profiles, occupation statistics, stochasticity, and their covariance. For the present proof-of-concept study, using the directly measured 1-halo contribution is the cleanest way to isolate the accuracy of the new 2-halo calculation.

  \subsection{Simulation and numerical implementation}\label{ssec:th.sim}
    The measurements used in this work are based on a gravity-only $N$-body simulation from the BACCO suite \citep{2004.06245}. These simulations have been extensively used for precision studies of large-scale structure \citep{2407.07949} and provide a controlled environment for testing halo-model predictions. The simulation considered here evolves $1536^3$ particles in a periodic box of side length $L=512\,h^{-1}{\rm Mpc}$, corresponding to a particle mass of $m_p\simeq3.2\times10^9\,h^{-1}M_\odot$. The gravitational evolution is performed using a modified version of the L-Gadget3 code \citep{0505010}. The simulation adopts a $\Lambda$CDM cosmology consistent with the Planck 2018 results \citep{1807.06209}, with parameters $\Omega_m=0.3096$, $\Omega_b=0.0490$, $\Omega_\Lambda=0.6904$, $h=0.6766$, $n_s=0.9665$, and $\sigma_8=0.8102$.

    To reduce the computational cost of the field-level measurements, we construct the CHEFT fields using a subsampled particle distribution, selecting one particle out of every $4^3$ in Lagrangian space. This yields a catalogue containing $1/64$ of the original particles. We have verified that this subsampling has a negligible impact on the large-scale power spectra and on the operator fields over the range of scales considered here. Halo-level quantities, including halo masses and positions, are computed from the full simulation data. The initial amplitudes of the simulation were fixed to the linear power spectrum in order to reduce cosmic variance on large scales \citep{1603.05253}.

    Haloes are identified using a Friends-of-Friends algorithm with linking length 0.2 and a minimum of 20 particles per halo. This gives haloes in the mass range $M\in[6.8\times10^{10},4.1\times10^{15}]\,h^{-1}M_\odot$. We divide this range into 100 logarithmic mass bins when carrying out halo-model sums. In all such sums we include an additional bin corresponding to field particles, i.e. particles not assigned to any resolved halo, which are treated as unresolved objects of mass $m_p$. This preserves the full mass budget of the simulation and makes the matter consistency tests in Section~\ref{sec:matter} meaningful. For this proof-of-concept study, we use only the $z=0$ snapshot. This is a conservative choice, since non-linear effects in the clustering of matter and haloes are stronger at $z=0$ than at earlier times. A full emulator for the CHEFT basis spectra and bias functions across cosmology and redshift is left for future work.


\section{Validation on the matter field}
\label{sec:matter}

Before exploring the CHEFT approach for general LSS tracers, it is important to verify that it provides a consistent description of the matter power spectrum. This constitutes a non-trivial internal validation of the framework, since all ingredients entering the model are measured directly from the same simulation. This test also allows us to quantify the level of accuracy expected from the finite simulation volume and from the numerical uncertainties associated with the measured ingredients, in particular the halo bias functions.

\subsection{Self-consistency of the model and bias normalisation constraints}
\label{ssec:matter.constraints}

For the matter overdensity field, the halo-model prediction reduces, in principle, to a consistency relation. If the decomposition into 1-halo and 2-halo contributions is complete, the sum of the measured 1-halo term and the CHEFT-based 2-halo term should reproduce the total matter power spectrum measured directly from the simulation.

Within the CHEFT framework, this requirement translates into a constraint on the mass-weighted bias integrals. Mass conservation implies that all bias-weighted contributions beyond the constant operator must vanish,
\begin{equation}
\label{eq:norm_bias}
I^i_m = \sum_n \frac{N_nM_n}{V_{\rm box}\bar\rho_m} b^{\mathcal{L}}_{i,n} =0,
\qquad i\neq 1.
\end{equation}
Here $N_n/V_{\rm box}$ is the number density of haloes in mass bin $n$, and $M_n/\bar\rho_m$ is the corresponding contribution to the fractional matter field. Physically, this condition reflects the fact that matter is an unbiased tracer of itself and should not carry higher-order bias contributions \citep{1511.02231}.

In practice, these constraints are only approximately satisfied. This is due to the finite volume of the simulation, the discretisation of the halo mass range, and the numerical uncertainties in the measured bias functions $b_i^{\mathcal L}(M)$. For the operators considered here we find
\begin{align}
I_m^\delta &= -0.01, \qquad
I_m^{\delta^2} = -0.016, \qquad
I_m^{s^2} = 0.004, \\ &
I_m^{\nabla^2\delta} =
-1.84\left[h\,{\rm Mpc}^{-1}\right]^2 .
\nonumber
\end{align}
The first three quantities are dimensionless and can be directly interpreted as small violations of the matter consistency relation. The Laplacian integral, however, carries units of $[h\,{\rm Mpc}^{-1}]^2$, and therefore its numerical value cannot be compared directly with the other constraints. For this reason, we do not use the absolute value of $\smash{I_m^{\nabla^2\delta}}$ as the criterion for keeping or rejecting this contribution. Instead, we assess the Laplacian term operationally, through its impact on the reconstructed matter power spectrum.

The Laplacian operator is particularly sensitive to this test because it depends on the smoothing prescription used to define the linear density field. In our implementation, two smoothing scales enter independently: the smoothing scale used in the probabilistic-bias measurement of $\smash{b_i^{\mathcal L}(M)}$, and the smoothing scale used to construct the operator spectra $\smash{P_{\rm CHEFT}^{ij}(k)}$. We adopt $k_s^{\rm PB}=0.3,h{\rm Mpc}^{-1}$ for the probabilistic-bias measurements, while the fiducial CHEFT operator spectra are constructed without additional smoothing. This mismatch is expected to affect higher-derivative operators more strongly than the lower-order density and tidal terms.

As shown in Fig.~\ref{fig:power_matter}, including the fixed probabilistic-bias measurement of $b_{\nabla^2\delta}^{\mathcal L}(M)$ together with the fiducial CHEFT Laplacian spectra spoils the percent-level matter reconstruction obtained with the lower-order operators. This indicates that, with the current smoothing prescriptions, the Laplacian contribution is not a stable prediction of the baseline model.

We therefore define our minimal baseline model, CHEFTmin, by excluding the Laplacian contribution and retaining the operator set
$\mathrm{CHEFTmin}=\{1,\delta_{\rm L},\delta_{\rm L}^2,s^2\}$.
Unless stated otherwise, all baseline predictions in the following sections refer to this CHEFTmin model.

This choice should not be interpreted as evidence that higher-derivative effects are physically irrelevant. Rather, it indicates that the Laplacian contribution must be treated consistently with the coarse-graining prescription. In Section~\ref{ssec:non_mater.extended}, we therefore introduce an extended model, CHEFText, in which the Laplacian contribution is restored as an effective calibrated term. This provides a controlled way of absorbing the unresolved smoothing dependence while preserving the simulation-driven nature of the approach.

To quantify the impact of the residual violations of Eq.~\eqref{eq:norm_bias}, we also study the effect of enforcing the normalisation condition on the reconstructed matter power spectrum. We impose this condition by introducing a compensating low-mass contribution, with effective mass $M_{\rm lm}$ and abundance $N_{\rm lm}$, such that
\begin{equation}
\sum_n\frac{N_n M_n}{V_{\rm box}}b_{i,n}^{\mathcal L}(M_n)+\frac{N_{\rm lm} M_{\rm lm}}{V_{\rm box}}b_i^{\mathcal L}(M_{\rm lm})=0.
\end{equation}
This procedure enforces the mass-conservation constraint while preserving the structure of the halo-model calculation, including the halo-profile contribution.

\subsection{Results}
\label{ssec:matter.res}

\begin{figure}
\includegraphics[width=0.98\columnwidth]{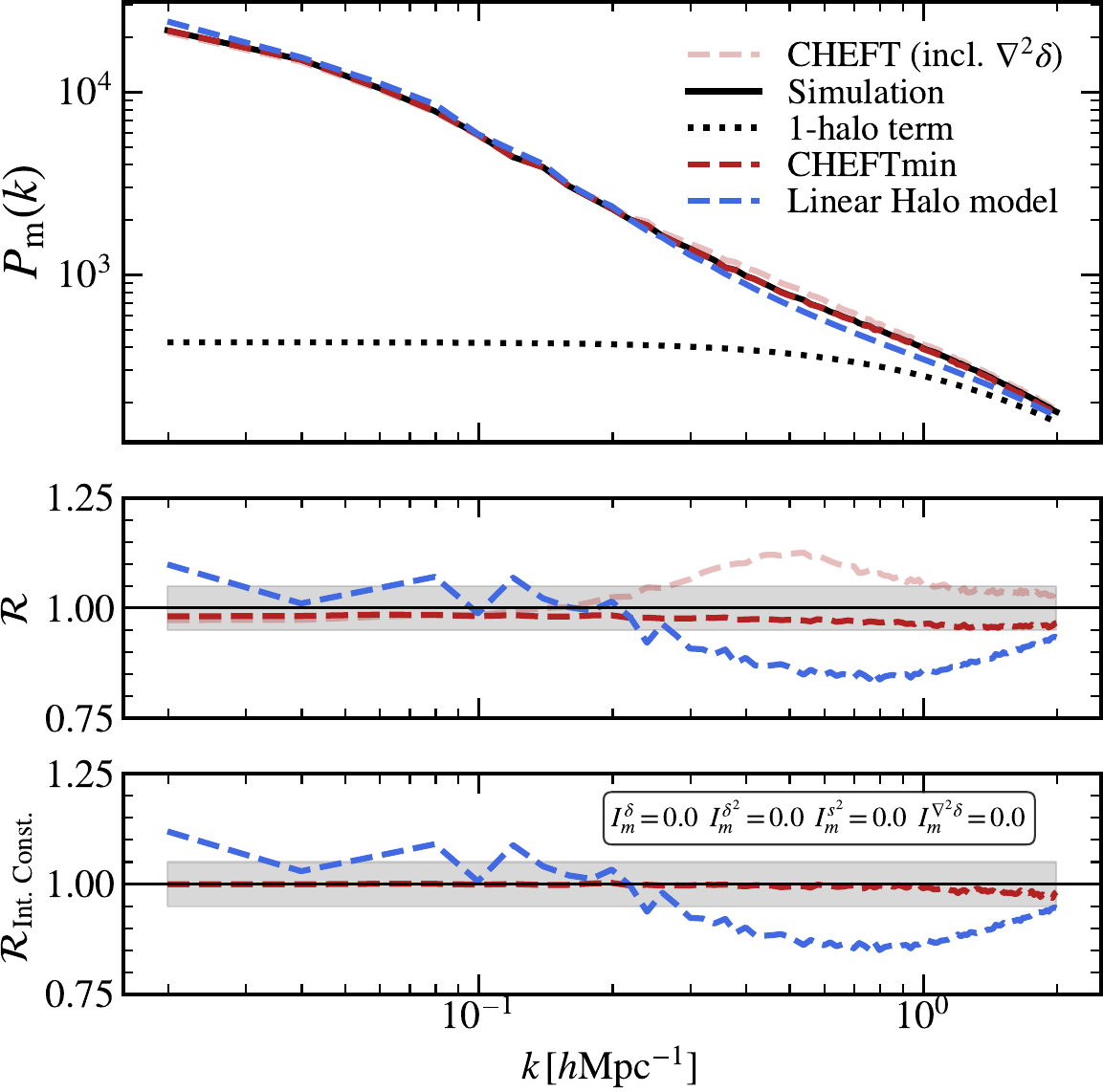}
\caption{Validation of the hybrid model on the matter power spectrum. The total power spectrum is reconstructed as the sum of the measured 1-halo term and the CHEFT-based 2-halo contribution. The top panel compares the reconstructed spectra with the simulation measurement, while the lower panels show the ratio residuals $\mathcal R$. The blue dashed line shows the standard linear-bias halo model. The red dashed line shows CHEFTmin, defined by the operator set ${1,\delta_{\rm L},\delta_{\rm L}^2,s^2}$, while the light-red dashed line shows the result of adding the fixed Laplacian contribution measured from the probabilistic-bias approach. Including this fixed Laplacian term worsens the matter reconstruction, illustrating the sensitivity of the higher-derivative contribution to the smoothing prescription. The lowest panel shows the effect of enforcing the mass-conservation constraints of Eq.~\eqref{eq:norm_bias}, which restores percent-level agreement down to $k\simeq1.5\,h\,{\rm Mpc}^{-1}$.}
\label{fig:power_matter}
\end{figure}

Figure~\ref{fig:power_matter} shows the matter power spectrum together with predictions from different flavours of the halo model. The standard linear-bias halo model displays deviations of order $\sim$\,20\% over the range $0.3\,h\,{\rm Mpc}^{-1}\lesssim k\lesssim2\,h\,{\rm Mpc}^{-1}$, corresponding to the transition between the 1-halo and 2-halo regimes.

The CHEFTmin prediction substantially reduces these deviations. Without enforcing the bias-normalisation constraints, the reconstructed matter power spectrum agrees with the simulation at the $2$--$5\%$ level. The remaining discrepancy is mainly driven by the small residual violation of the linear-bias normalisation condition, while the quadratic and tidal contributions are subdominant. Adding the fixed Laplacian contribution does not improve this reconstruction; instead, it introduces a coherent scale-dependent deviation at intermediate scales. This motivates excluding the fixed Laplacian term from the minimal baseline model.

After imposing the mass-conservation constraints, the CHEFTmin reconstruction reaches percent-level agreement down to $k\simeq1.5\,h\,{\rm Mpc}^{-1}$. Since an analogous constraint cannot be imposed in a model-independent way for general weighted tracers, the residual accuracy of the unconstrained CHEFTmin matter reconstruction provides a useful estimate of the precision floor associated with the finite simulation volume and the measured bias functions. We therefore use the few-percent level reached in this test as the target accuracy for the weighted-tracer results presented below.

In spite of these residual errors, the CHEFT-based 2-halo term largely removes the transition-scale inaccuracies of the linear halo model. This validates the internal consistency of the collapsed-operator construction and supports its application to the more general weighted observables considered in the next section.

      \begin{figure*}
	    \includegraphics[width=\textwidth]{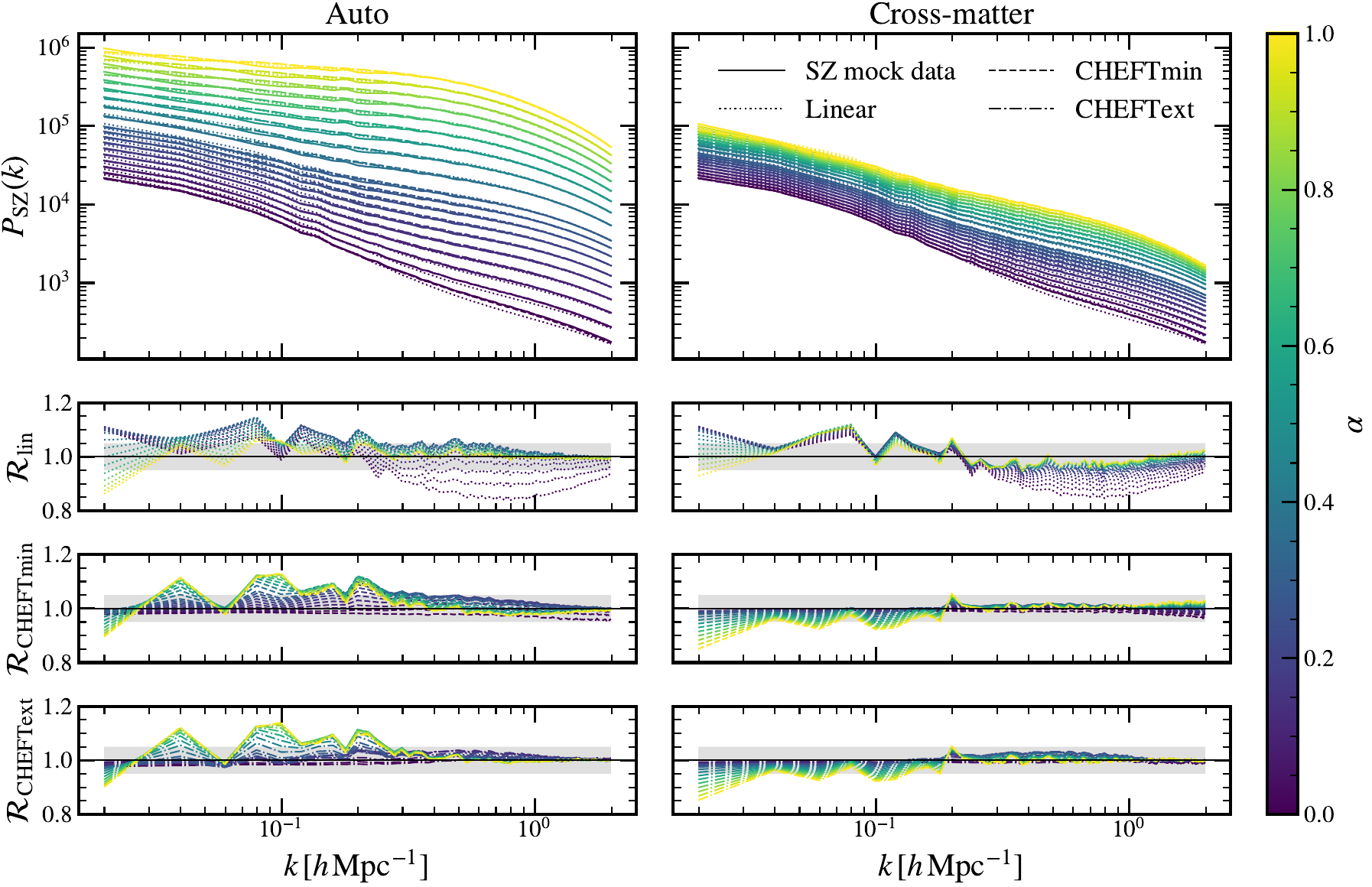}
        \caption{Power spectra for SZ-like weighting with different values of the exponent $\alpha$. Solid lines show the simulation results, while dashed and dotted lines correspond to the linear halo model and the CHEFT predictions, respectively. The lower panels display the ratio residuals $\mathcal{R}$. The CHEFTmin model significantly improves the description of the transition regime for low values of $\alpha$, where the signal receives contributions from a broad range of halo masses. For larger $\alpha$, the clustering becomes dominated by massive haloes and the 1-halo term, reducing the difference between models. The cross-correlation with matter shows consistent percent-level improvements across all cases. The extension to CHEFTmin, CHEFText, gains a few percent precision, leaving the accuracy of the model at around 3\%.}\label{fig:Malpha}
      \end{figure*}
      
\section{Results for non-matter tracers}\label{sec:non_matter}
  We now assess the performance of the hybrid EFT-halo model on a set of mock tracers of the LSS constructed by applying weights to the simulation particles dependent on halo mass (see Section \ref{ssec:th.wM} for details), mimicking the behaviour of real-world probes. These tests probe the ability of the model to reproduce clustering statistics when different halo mass ranges contribute with varying importance.

  \subsection{Performance of the minimal CHEFT model}\label{ssec:non_mater.minimal}

    We begin by evaluating the predictive performance of the minimal CHEFT-based 2-halo term, including only the contributions from linear, quadratic, and tidal bias, without introducing any additional free parameter. In all cases, we combine the CHEFT 2-halo contribution with the 1-halo term directly measured from the simulation.

    \subsubsection{SZ-like weighting}
      The results for the SZ-like weighting, where the weights are proportional to $M^\alpha$, are shown in the upper and middle panels of Fig.~\ref{fig:Malpha}. Results are shown for the auto-correlation of these mock tracers and for their cross-correlation with the matter overdensity. In addition to the power spectra measured in the simulations, we show the halo model predictions using the linear bias model (dotted lines) and the minimal CHEFT (CHEFTmin) approach (dashed lines) to describe the 2-halo contribution. Focusing first on the auto-correlation, we find that, for low values of the exponent $\alpha$, the CHEFT model significantly improves the standard linear halo model on the scales of the transition between the 1-halo and the 2-halo regimes. As $\alpha$ increases, the auto-spectrum becomes increasingly dominated by massive haloes, and the 1-halo term contributes a larger fraction of the total power. In this regime, both the linear and CHEFT models provide similar levels of accuracy of around 5\% since the intermediate scales become dominated by the 1-halo term. Note also that the large scales become significantly noisier for large values of $\alpha$, caused by the small number of high-mass haloes, which increases the level of cosmic variance in the auto-power spectrum measurements. This prevents us from assessing the accuracy of the CHEFT prediction below the $\sim$\,5\% threshold in this regime.

      In the case of the cross-power spectrum with the matter field, the CHEFT model consistently outperforms the linear approximation, achieving agreement at the $\sim$\,2\% level down to scales of $k \sim 1.5\,h\,\mathrm{Mpc}^{-1}$. This is a significant improvement from the $\sim$\,20\% precision of the linear bias model.

  \subsubsection{CIB and HOD-like weighting}
    \begin{figure*}
      \includegraphics[width=0.49\textwidth]{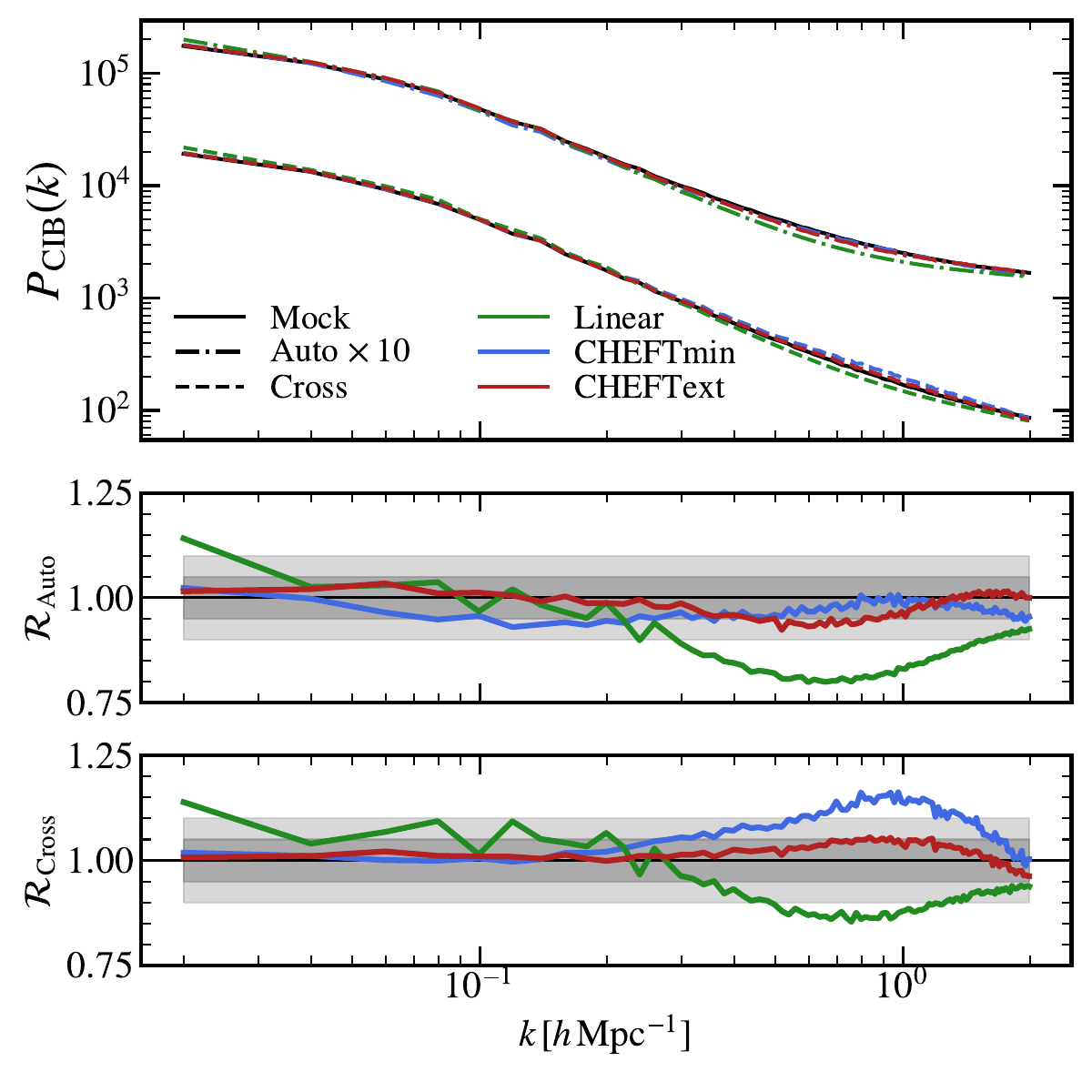}
	   \includegraphics[width=0.49\textwidth]{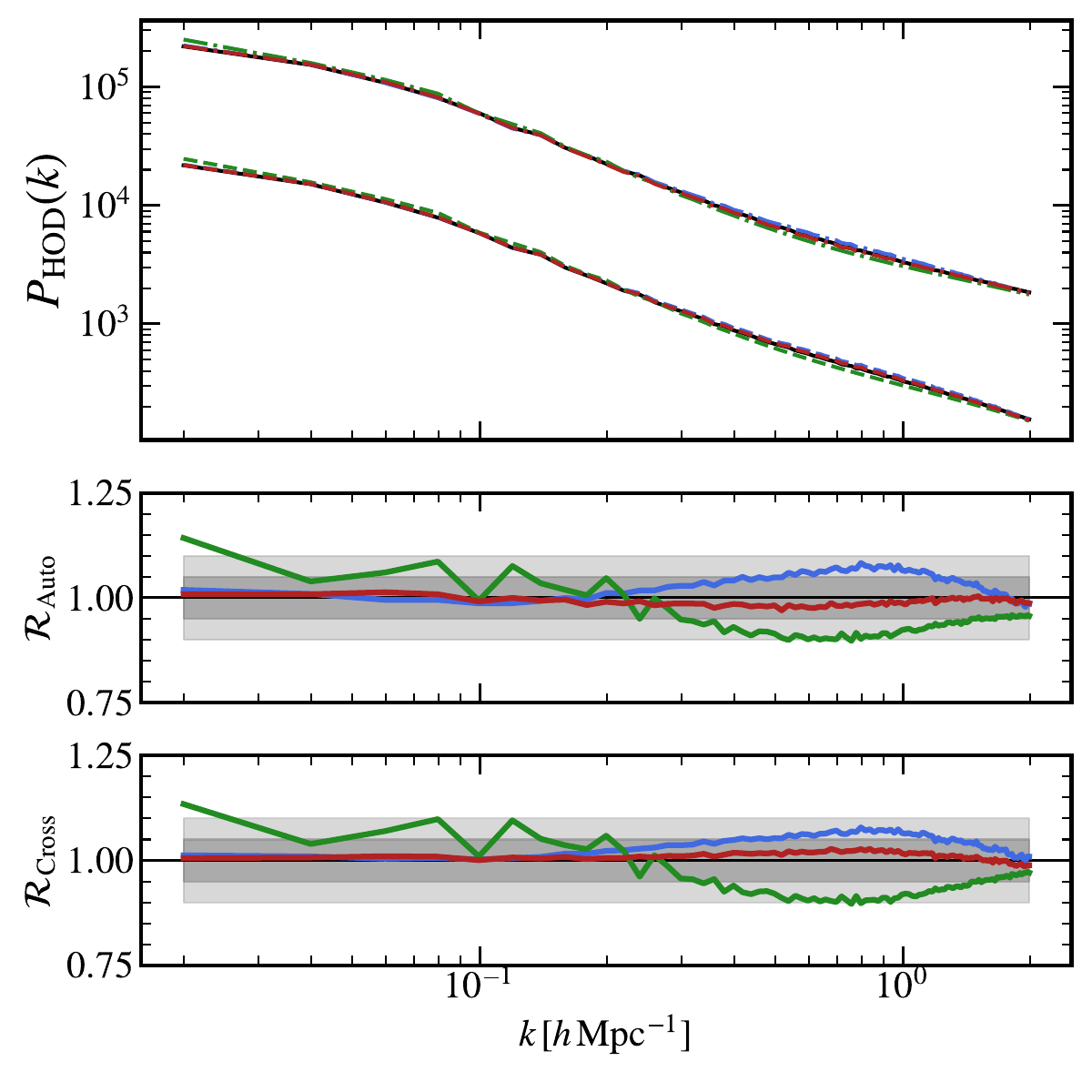}
      \caption{The same as Fig.~\ref{fig:Malpha}, but for the CIB-like weighting (left panels) and HOD-like weighting (right panels). In both cases, the minimal CHEFT model, CHEFTmin, improves over the standard linear-bias halo model but exhibits coherent residuals at intermediate scales. These deviations are particularly visible in the cross-correlations with matter, indicating that the second-order CHEFT basis does not fully capture the mass-dependent halo-halo correlations relevant for these weighting schemes. The extended model, CHEFText, in which the Laplacian contribution is promoted to an effective parameter, significantly reduces these residuals. For the HOD-like weighting, CHEFText achieves few-percent agreement for both the auto- and cross-spectra across most of the range shown. For the CIB-like weighting, the improvement is also substantial, although some residual scale dependence remains. As discussed in Appendix~\ref{Appendix:Phh_fit}, the limitations of CHEFTmin are most pronounced for observables receiving comparable contributions from a broad range of halo masses, where inaccuracies in the full $P_{hh}(k|M_1,M_2)$ are less efficiently averaged out.}
\label{fig:CIBHOD}

    \end{figure*}
    The performance of the model for the CIB-like weighting is shown in the left panel of Fig.~\ref{fig:CIBHOD}. In this case, the minimal CHEFT prediction exhibits more noticeable deviations from the simulation results, particularly at intermediate scales. These discrepancies are larger than those observed in the SZ case and indicate that the model does not fully capture the relevant clustering physics for this weighting scheme. Although the CHEFT prediction does provide an improvement over the linear model (e.g. from $\gtrsim$\,20\% to $\sim$\,10\% in the case of the auto-spectrum), the performance is significantly poorer than observed in the case of SZ-like weights.

    The results for the HOD-like weighting are presented in the right panel of Fig.~\ref{fig:CIBHOD}. The performance of the minimal CHEFT prediction is qualitatively similar to the CIB case, with the CHEFT model providing an overall improvement over the linear prediction, but failing to reach the same level of precision observed for the power spectrum of the matter overdensity and of SZ-like tracers. The similarity between these two cases is not entirely surprising, as the weighting schemes used in both cases suppress the contribution from low-mass haloes, and enhances the contribution from intermediate-mass haloes ($M\sim10^{11-13}\,h^{-1}M_\odot$; see Fig. \ref{fig:weights}).
    A failure to accurately reproduce the clustering properties of haloes in this mass range can then affect the performance of the CHEFT prediction significantly. We discuss these limitations in the next section.

\subsection{Limitations of the CHEFT expansion}\label{ssec:non_mater.limitations}
  
    The deviations observed for the CIB- and HOD-like weights indicate that the remaining inaccuracies of the minimal CHEFT model are not simply associated with a particular observable, but with the description of the mass-dependent halo-halo power spectrum itself. At fixed wavenumber, $P_{hh}(k|M_1,M_2)$ should be regarded as a matrix in halo-mass space. As discussed in Section~\ref{sec:th}, the standard linear-bias halo model approximates this matrix by a single outer product, and therefore assumes perfect coherence between all halo populations. The CHEFT expansion relaxes this structure, with its full factorisation in halo mass, by allowing different halo masses to project onto several non-linear Lagrangian operator fields. However, the second-order CHEFT basis is still a finite deterministic expansion, and the off-diagonal structure of this matrix, i.e. the cross-correlations between different halo masses, provide a more stringent test than the auto-spectra of individual mass bins.

    The tests presented in Appendix~\ref{Appendix:Phh_fit} show precisely this behaviour. The CHEFT expansion can reproduce the auto-spectra $P_{hh}(k|M,M)$ of individual halo mass bins with good accuracy, but the same fitted operator basis does not provide an equally accurate description of the full $P_{hh}(k|M_1,M_2)$ matrix, especially for cross-correlations between well separated halo masses. This distinction is important: the auto-spectra mainly test the clustering amplitude assigned to each halo population, whereas the cross-spectra test whether the model captures the coherence between different halo populations. The second-order Lagrangian operator basis therefore captures a large fraction of the non-linear halo clustering signal, but does not fully describe the cross-mass coherence structure of the halo field.

    This explains why the impact of the missing off-diagonal structure depends strongly on the tracer weighting. If an observable is dominated by a single effective halo population, or if its mass weighting projects mostly onto combinations of $P_{hh}(k|M_1,M_2)$ that are already well captured by the CHEFT basis, the residual cross-mass errors have little impact on the final integrated spectrum. By contrast, observables that receive comparable contributions from several intermediate halo masses are more sensitive to the detailed off-diagonal structure of the halo--halo matrix. This is the case for the CIB- and HOD-like weights considered here, which emphasise intermediate halo masses and therefore provide a stringent test of the cross-mass coherence of the CHEFT expansion.

    The results of Appendix~\ref{Appendix:Phh_fit} also indicate that this behaviour is not removed simply by increasing the freedom of the fitted bias parameters within the CHEFTmin basis, nor by the modest extensions of the operator basis explored there. Instead, it points to missing ingredients in the present description of halo clustering, with natural candidates including higher-derivative contributions, halo exclusion, stochasticity, and possible dependence on secondary halo properties. We leave a systematic treatment of these effects for future work. In the next section, we show that a simple effective treatment of the Laplacian contribution already absorbs a significant fraction of the residual mismatch in the observables considered here.

  \subsection{Extended model: effective Laplacian rescaling}\label{ssec:non_mater.extended}
    The discussion above identifies the incomplete description of the mass-dependent halo-halo matrix as the main limitation of CHEFTmin. Rather than attempting to model all missing effects separately, we now ask whether a compact effective correction can absorb the leading residual contribution in the weighted observables
    considered here.
    
    The natural candidate is the higher-derivative Laplacian operator. As discussed in Section~\ref{sec:matter}, the fixed probabilistic-bias prediction for $b^L_{\nabla^2\delta}(M)$ is sensitive to the smoothing prescription and does not provide a stable contribution to the matter reconstruction. We therefore define an extended model, CHEFText, in which the Laplacian bias function is replaced by a smooth effective rescaling,
    \begin{equation}
      b^{\mathcal{L}}_{\nabla^2\delta}(M)\rightarrow b^{\rm eff}_{\nabla^2\delta}(M) = \alpha(M)b^{\mathcal{L},s}_{\nabla^2\delta}(M),
      \label{eq:beff_laplacian_main}
    \end{equation}
    where $\smash{b^{\mathcal{L},s}_{\nabla^2\delta}(M)}$ is a smoothed version of the
    probabilistic-bias measurement and $\alpha(M)$ is taken to be a
    linear function of $\log_{10}M$.
    
    An important result is that this calibration appears to be largely
    probe independent. Fitting the same parameters separately to each
    tracer changes the resulting predictions only at the percent level, as
    shown in Appendix~\ref{Appendix:Phh_fit}. Thus CHEFText should
    not be interpreted as adding arbitrary tracer-dependent freedom, but
    as an effective, simulation-calibrated higher-derivative correction
    that can be applied consistently across the different mass-weighted
    observables considered here.
    
    The performance of CHEFText is shown by the dashed-dotted and red curves in
    Figs~\ref{fig:Malpha} and~\ref{fig:CIBHOD}. For the
    SZ-like and HOD-like weights, the agreement with the simulation
    improves to the level of $\sim 3\%$ over most of the range
    $0.2\lesssim k\lesssim 1.5\,h{\,\rm Mpc}^{-1}$. For the CIB-like
    weighting, the accuracy improves to the $\sim$\,5\% level, although
    some residual scale dependence remains. This demonstrates that a
    single effective Laplacian calibration captures much of the residual
    mismatch without changing the overall structure of the hybrid
    halo-model construction.

\section{Summary and Discussion}
\label{sec:disc}
  The results presented in this work demonstrate that a hybrid EFT--halo model can substantially improve the description of the 2-halo contribution relative to the standard halo model implementation based on linear halo bias. The origin of this improvement is straightforward: the traditional approximation compresses the halo--halo clustering signal into a linear-bias rescaling of the linear matter power spectrum, whereas the CHEFT construction replaces this with a basis of non-linear Lagrangian operators measured directly from simulations.

  The collapse step in the construction of this EFT-inspired basis is essential to make it compatible with the halo model. By collapsing particles belonging to the same halo to a common halo centre, the operator fields remove intra-halo structure by construction, while retaining the inter-halo correlations relevant for the 2-halo term. This allows us to maintain the modular structure of the halo model, including the integration over halo mass and halo profiles, while replacing the linear treatment of halo clustering with a simulation-calibrated non-linear expansion.

  At the matter level, this construction passes a non-trivial consistency test: combining the measured 1-halo term with the CHEFT-based 2-halo contribution recovers the matter power spectrum at percent-level accuracy across the transition regime. For the weighted observables considered in this work, the minimal CHEFT model typically achieves $\sim$\,5\% accuracy.
  This represents a significant improvement over the standard linear-bias halo model, and shows that a relatively compact operator basis can capture a large fraction of the relevant non-linear clustering physics. 
  
  The main limitation of CHEFTmin is its incomplete description of the off-diagonal, cross-mass structure of $P_{hh}(k|M_1,M_2)$. Appendix~\ref{Appendix:Phh_fit} shows that the basis can reproduce the clustering amplitudes of individual halo mass bins, but does not fully recover the coherence between different halo populations. This limitation is most visible for the CIB- and HOD-like weights, which are sensitive to several intermediate halo masses.

  We have shown that the Laplacian operator can play a special role in this context. Its contribution depends on the smoothing scale used to define the linear density field, since a higher-derivative term is naturally sensitive to the details of the coarse-graining prescription. We therefore introduced the CHEFText model, in which the Laplacian contribution is treated as an effective mass-dependent correction calibrated from simulations. This minimal extension improves the agreement with the measured weighted power spectra to the $\sim$\,$3-5\%$ level for all tracers explored here. This improvement over the linear-based halo model could be particularly relevant in multi-probe analyses that combine traditional LSS observables (e.g. galaxy clustering and weak lensing) with baryonic probes such as the Sunyaev--Zeldovich effects, the CIB, or diffuse X-ray maps \citep[e.g.][]{2109.04458,2204.01649,2209.05472,2309.11129,2410.22397,2412.12081,2506.07432,2509.17539,2509.03458,2505.20413}.
  
  These results motivate future work in the following directions:
  \begin{itemize}
    \item The construction of emulators for the CHEFT operator spectra and their associated bias functions. This would require the use of larger simulations, but could reuse the simulation suites and techniques employed in existing implementations of the HEFT approach \citep{2101.12187,2303.09762}.
    \item A more detailed study of the impact of halo definition and halo finding. We have used FoF haloes, and the measured CHEFT spectra, halo profiles, bias functions, and 1-halo contributions are tied to this halo definition. Exploring alternative halo definitions should give an insight into the robustness of the model. 
    Other limitations, such as the mass resolution of the simulation, the treatment of low-mass haloes, and the precision of the measured halo bias relations, should also be quantified in more detail. 
    \item We have shown that the CHEFText model is not a perfect match to the halo-halo power spectrum on intermediate and small scales. Instead, it is a compact effective description of the leading impact of non-linear bias. A more complete treatment would require a controlled modelling of higher-derivative bias, and of stochastic bias (e.g. halo exclusion effects) in the cross-spectrum between haloes of different masses.
    \item Finally, having improved the modelling of the 1-halo to 2-halo transition regime, the main remaining source of theoretical uncertainty lies in the halo profiles associated with relevant cosmological observables, including weak lensing, CMB secondary anisotropies etc. Besides capturing the average behaviour of the appropriate astrophysical properties (e.g. matter density, thermal gas pressure), these models should also account for the stochasticity of these quantities and the covariance between them, which directly impact the 1-halo contribution to their power spectra.
  \end{itemize}
  A full implementation of these ingredients would provide a more accurate and robust halo-model framework for joint analyses of multiple large-scale structure tracers.

\section*{Acknowledgements}
MPI would like to thank Martin White and Caroline Guandalin for useful discussions. DA acknowledges support from the Beecroft Trust.
MZ acknowledges that the project that gave rise to these results received the support of a fellowship from the `La Caixa' Foundation (ID 100010434); the fellowship code is LCF/BQ/PI25/12100027.

\section*{Data Availability}
  The data underlying this article will be shared on reasonable request to the corresponding author.



\bibliographystyle{mnras}
\bibliography{main} 



\appendix

\section{CHEFT basis terms}
\label{Appendix:basis_terms}  
\begin{figure}  \includegraphics[width=0.98\columnwidth]{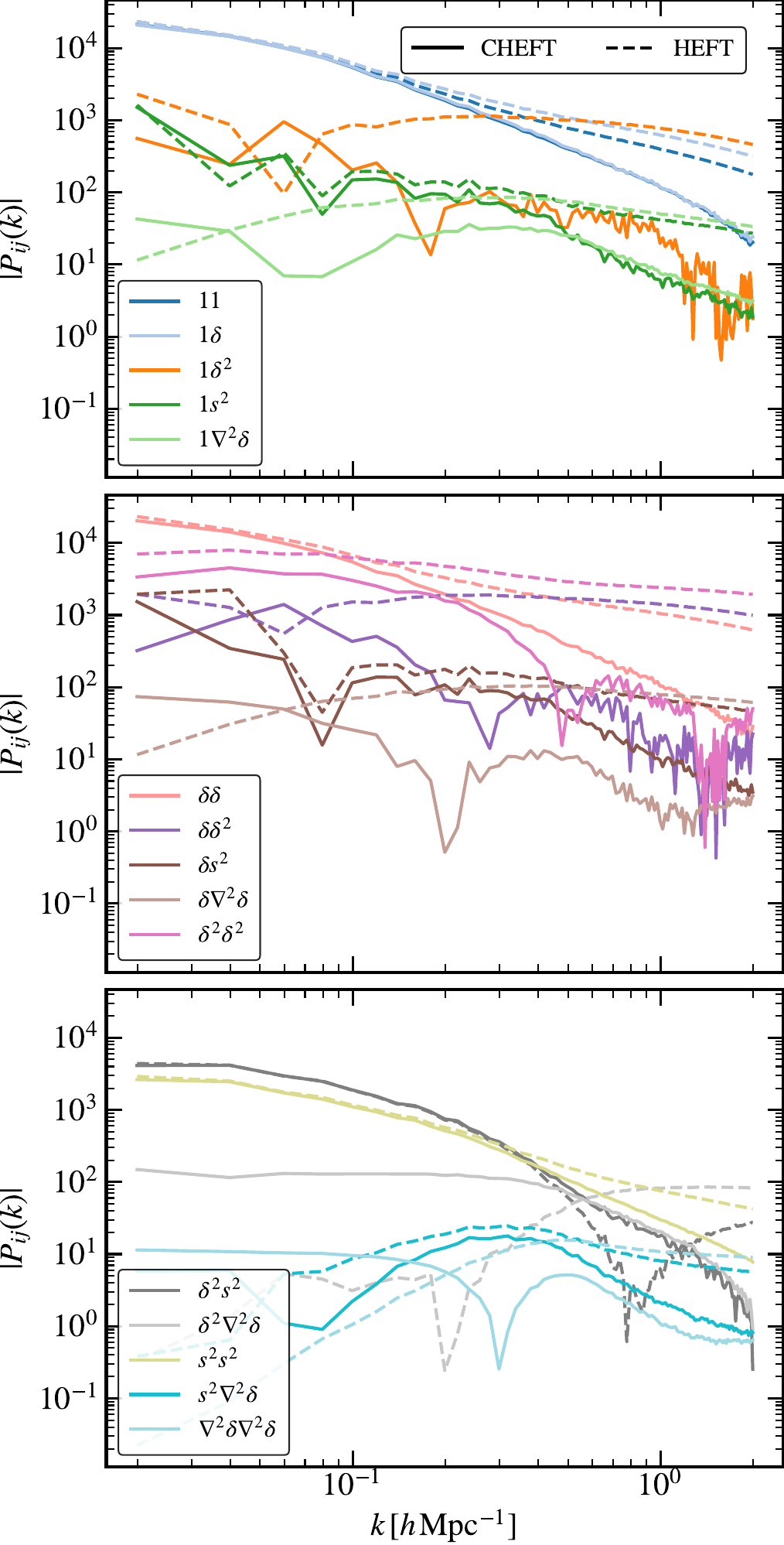}
  \caption{Comparison between the standard HEFT basis spectra and the corresponding collapsed CHEFT basis spectra. Dashed lines show the HEFT spectra, while solid lines show the CHEFT spectra obtained after collapsing particles to their halo centre. The collapse procedure removes intra-halo contributions from each weighted field, leading to a suppression of small-scale power and modifying the relative shapes of several operator spectra. All spectra are computed using a smoothing scale of $k_s=0.75\,h\,{\rm Mpc}^{-1}$. }
  \label{fig:basis_terms}
\end{figure}

In this appendix, we compare the operator power spectra entering the CHEFT basis with the corresponding spectra obtained from the standard, uncollapsed HEFT fields. This comparison is useful for illustrating the direct impact of the collapse procedure on the individual basis terms. The results are shown in Fig.~\ref{fig:basis_terms}.
The CHEFT spectra generally approach the corresponding HEFT spectra on large scales for the lower-order operators, as expected since the collapse modifies only intra-halo structure. The agreement is less clean for terms involving the Laplacian operator. These terms are particularly sensitive to smoothing and to the weighted shot-noise subtraction required for the collapsed fields.

The clearest example is the $11$ contribution. In the standard HEFT construction, this term corresponds to the matter density field and therefore contains both intra-halo and inter-halo correlations. In the CHEFT procedure, particles belonging to the same halo are moved to the halo centre, removing the contribution associated with the internal halo density profile. As expected, this leads to a suppression of power on small scales.

  A similar comparison can be made for the $1\delta$ term. In the uncollapsed HEFT basis, this spectrum differs noticeably from the $11$ contribution. Once the weighted intra-halo contribution is removed through the CHEFT construction, however, the $11$ and $1\delta$ spectra become much more similar in shape. This suggests that a significant part of the difference between these two terms in the standard HEFT basis is sourced by correlations internal to haloes. The remaining operator combinations show analogous behaviour.

  For this comparison we use a smoothing scale of $k_s=0.75\,h\,{\rm Mpc}^{-1}$, chosen to facilitate comparison with previous HEFT measurements in the literature \citep[e.g.][]{2101.11014,2101.12187,2207.06437,2303.09762}. The spectra shown here are noisier than those presented in those works for two main reasons. First, although the simulation used here is fixed and paired, its volume is smaller, leading to a larger residual contribution from cosmic variance. Second, we do not replace the largest-scale modes by perturbation-theory predictions, and therefore retain the full simulation noise on those scales.

  The CHEFT spectra are also generally noisier than the corresponding HEFT spectra. This is partly a consequence of the collapse operation itself, which depends on the halo catalogue and therefore inherits noise associated with the halo-finding procedure. In addition, the collapsed fields require a shot-noise subtraction. We estimate this contribution using randomised catalogues with the same number density and weights as the collapsed fields, and averaging over 10 random catalogues. Increasing the number of random realisations would reduce 
  the uncertainty in this process.

  On sufficiently large scales, the collapsed CHEFT spectra are expected to approach their HEFT counterparts for operator combinations whose stochastic contribution is subdominant, since the collapse only modifies the internal distribution of particles within haloes. This behaviour is visible for several of the lower-order operators in Fig.~\ref{fig:basis_terms}. However, the agreement is not exact term by term. In particular, spectra involving higher-derivative operators, such as $\nabla^2\delta$, are more sensitive to the smoothing prescription and to the subtraction of the weighted shot-noise contribution associated with the collapsed halo catalogue. The random-catalogue subtraction removes the leading Poisson contribution, but residual non-Poisson or scale-dependent stochastic terms can remain. We therefore use Fig.~\ref{fig:basis_terms} primarily as a diagnostic of the impact of the collapse procedure, rather than as a precision test of the low-$k$ perturbative limit of each individual operator spectrum.

\section{CHEFT and the halo-halo power spectrum}\label{Appendix:Phh_fit}
  In this appendix, we investigate whether the CHEFT framework can be extended to accurately reproduce the full halo-halo power spectrum $P_{hh}(k|M_1,M_2)$ by increasing the flexibility of the model. This analysis is designed to test whether the discrepancies identified in Section~\ref{ssec:non_mater.limitations} arise from limitations in the bias parametrisation, the choice of operator basis, or other modelling assumptions.

  The purpose of this test is not only to determine whether the CHEFT basis can fit the clustering amplitude of each halo mass bin, but also whether it can predict the coherence between different mass bins. These are distinct requirements. The auto-spectrum $P_{hh}(k|M,M)$ fixes the norm of the model vector associated with mass $M$, while the cross-spectrum $P_{hh}(k|M_1,M_2)$ tests the angle, or overlap, between the model vectors associated with two different halo masses. A finite deterministic bias expansion can therefore fit all auto-spectra reasonably well while still failing to reproduce the full cross-mass matrix. The latter is the more stringent test, and is the one most relevant for weighted observables whose 2-halo terms involve double integrals over halo mass.

  \subsection{Methodology}
    We consider halo masses in the range $M \in [6.8 \times 10^{10},\, 4.1 \times 10^{15}]\,h^{-1}M_\odot$, divided into 100 logarithmic bins, and evaluate the performance of different model configurations in reproducing the full two-dimensional function $P_{hh}(k|M_1,M_2)$.

    Our approach consists of two steps. First, we fit the halo auto-spectra $P_{hh}(k|M)$ and the cross-spectra with matter $P_{hm}(k|M)$ for each mass bin, allowing the bias parameters to be free and extending CHEFTmin to the model variations that we explain in the following section. Second, we use these fitted parameters to predict the full $P_{hh}(k|M_1,M_2)$, and evaluate the corresponding goodness-of-fit. This procedure allows us to disentangle the ability of the model to describe individual mass bins from its ability to capture correlations across different halo masses and circumvents the technical problem of fitting a 101$\times$101 function (100 halo mass bins plus 1 bin for field particles).

    For the fitting procedure we assume a Gaussian covariance for each spectrum independently (see e.g. \citealt{1610.06585}). This covariance is diagonal and can be written in terms of the power spectrum amplitude as
    \begin{equation}
    {\rm Cov}\!\left[P_{hh}(k_a),P_{hh}(k_b)\right]
    =
    \delta_{ab}\,
    \frac{2\,P_{hh}^2(k_a)}{N_{\rm modes}(k_a)},
    \end{equation}
    at the auto-spectrum level, while for the cross-bin terms we can write
    \begin{equation}
    {\rm Cov}\!\left[P_{hh^{\prime}}(k_a),P_{hh^{\prime}}(k_b)\right]
    =
    \delta_{ab}\,
    \frac{P_{hh}(k_a)P_{h^{\prime}h^{\prime}}(k_a)+P_{hh^{\prime}}^2(k_a)}
    {N_{\rm modes}(k_a)}.
    \end{equation}
    Note that, since we are only fitting to the auto-spectra and their correlation with matter, in the fitting procedure $h^{\prime}=m$. However, we leave the equation general because we later show the associated errorbars. The fit is found by minimising the $chi^2$ distance defined as 
    \begin{equation}
        \chi^2=\left( P_{hh}-P_{\rm C/HEFT} \right)^{\rm T}{\rm Cov}^{-1}\left( P_{hh}-P_{\rm C/HEFT} \right) ,
        \end{equation}    
    through the \texttt{Python} package \texttt{SciPy}\footnote{\url{http://www.scipy.org/}} \citep{1907.10121} with the Limited-memory BFGS method.

  \subsection{Model variations}
    We explore a range of extensions to the baseline CHEFTmin model:
    \begin{itemize}
      \item \textbf{Noise modelling}: We include scale-dependent noise terms in the auto-spectra to account for deviations from Poisson statistics and halo exclusion effects. We consider the following functional forms:
      \begin{align}
        &P^{\epsilon}_1(k)=\frac{1}{\bar{n}}(\epsilon_1 + \epsilon_2 k^2),\\
        &P^\epsilon_2(k)=\frac{1}{\bar{n}}(\epsilon_1 + \epsilon_2 k^2 + \epsilon_3 k^4),\\
        &P^\epsilon_3(k) = \frac{1}{\bar{n}}\,\epsilon_1 e^{-k^2 \epsilon_2}.
      \end{align}

      \item \textbf{Smoothing scale}: We vary the smoothing scale used to construct the operator fields over the range $k_{\rm s} = [0.75,\, 1.0,\, 2.0,\, 10.0,\, 100.0]\,h\,\mathrm{Mpc}^{-1}$.

      \item \textbf{Operator basis}: We test whether extending the CHEFT expansion to third order improves the reconstruction of $P_{hh}(k|M_1,M_2)$. In addition to the second-order operators, we include the cubic density field $\delta_{\rm L}^3$, the raw cubic tidal invariant $G_3$, and the mixed operator $\delta_{\rm L} G_2$, where $G_2$ and $G_3$ are constructed from the tidal tensor of the linear density field \citep{1611.09787}. Specifically, defining the tidal tensor as
      \begin{equation}
        T_{ij}(\mathbf{q}) \equiv \partial_i \partial_j \Phi(\mathbf{q}),
        \qquad \nabla^2 \Phi(\mathbf{q}) = \delta_{\rm L}(\mathbf{q}),
      \end{equation}
      the tidal quadratic and cubic operators are given by
      \begin{align}
        &G_2(\mathbf{q}) = \sum_{i,j} T_{ij}(\mathbf{q})\,T_{ij}(\mathbf{q})\\
        &G_3(\mathbf{q}) = \sum_{i,j,k} T_{ij}(\mathbf{q})\,T_{jk}(\mathbf{q})\,T_{ki}(\mathbf{q}),
      \end{align}
      and the mixed operator is defined as
      \begin{equation}
        \delta_{\rm L} G_2(\mathbf{q}) = \delta_{\rm L}(\mathbf{q})\,G_2(\mathbf{q}).
      \end{equation}

      In our implementation, these operators are constructed directly from the linear density field without applying Galilean transformations, and are normal-ordered by subtracting their mean values when required. We refer to this extension as the third-order (TO) basis, in contrast with the second-order (SO) model used in the main analysis.

      \item \textbf{Laplacian treatment}: We test both fixing the Laplacian contribution to zero and allowing it to vary freely together with the rest of the bias parameters.

      \item \textbf{Basis construction}: We compare the collapsed CHEFT fields with the standard (non-collapsed) HEFT fields, despite the latter including 1-halo contributions. We do this test for consistency.
    \end{itemize}

    For each configuration, we evaluate the reduced $\chi^2$ for both the auto-spectra and the full $P_{hh}(k|M_1,M_2)$ for the fitted case and for the Probabilistic Bias case.

    \begin{figure}
	   \includegraphics[width=0.98\columnwidth]{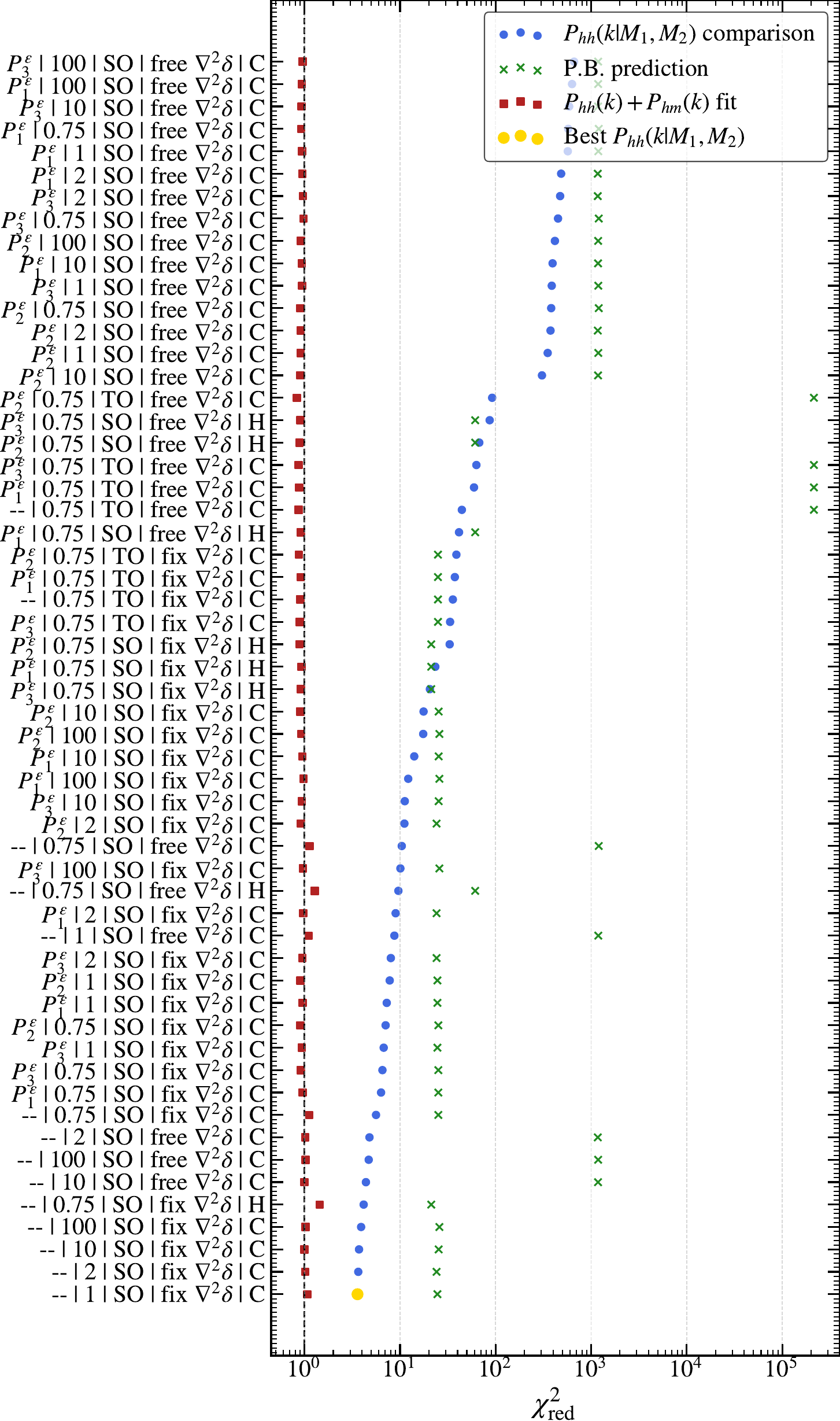}
      \caption{Reduced $\chi^2$ values for different model configurations tested in Appendix~\ref{Appendix:Phh_fit}. Each point corresponds to a specific combination of smoothing scale, noise model, operator basis, and Laplacian treatment. Results are shown separately for the auto-spectra and the full two-dimensional halo–halo power spectrum. While many configurations achieve good fits to the auto-spectra ($\chi^2 \sim 1$), none provide an acceptable description of the full $P_{\rm hh}(k|M_1,M_2)$, indicating a fundamental limitation of the operator expansion. The labels are built from the combination \{ Noise Model | Smoothing Scale | Operator Basis | Laplacian Treatment | Basis Construction \}. If the symbol -- is shown, this means that no Noise Model was included in the fitting procedure. }
      \label{fig:chi2_Phh}
    \end{figure}

\subsection{Results}

The results of this analysis are summarised in Fig.~\ref{fig:chi2_Phh}. We find that:
\begin{itemize}
  \item Most configurations provide good fits to the auto-spectra $P_{hh}(k|M,M)$, with reduced $\chi^2 \sim 1$.
  \item The predictions obtained using the probabilistic-bias relations alone perform poorly when evaluated on the full $P_{hh}(k|M_1,M_2)$ matrix.
  \item Allowing the bias parameters to vary improves the fit significantly, especially for the amplitudes of the individual spectra, but does not lead to an acceptable description of the full two-dimensional halo-halo power spectrum.
  \item None of the explored configurations achieve reduced $\chi^2 \sim 1$ for the full $P_{hh}(k|M_1,M_2)$ matrix, with the largest discrepancies appearing in cross-correlations between well separated halo masses.
\end{itemize}

Among the configurations explored, we find that the best-performing model corresponds to a setup with no additional noise parameters, a smoothing scale of $k_{\rm s} = 1\,h\,\mathrm{Mpc}^{-1}$ for the operator fields, the use of the collapsed CHEFT basis, and a second-order (SO) expansion, while fixing the Laplacian contribution to zero. This configuration yields the lowest $\chi^2$ values for the full $P_{hh}(k|M_1,M_2)$ among the models considered.

Many alternative configurations provide comparable fits to the auto-spectra, indicating a significant degree of degeneracy in the modelling choices when only individual halo mass bins are considered. In particular, variations in the smoothing scale, inclusion of higher-order operators, and different noise prescriptions can compensate each other at the level of $P_{hh}(k|M,M)$. The full matrix $P_{hh}(k|M_1,M_2)$ is therefore a more stringent diagnostic

\begin{figure}
\includegraphics[width=0.98\columnwidth]{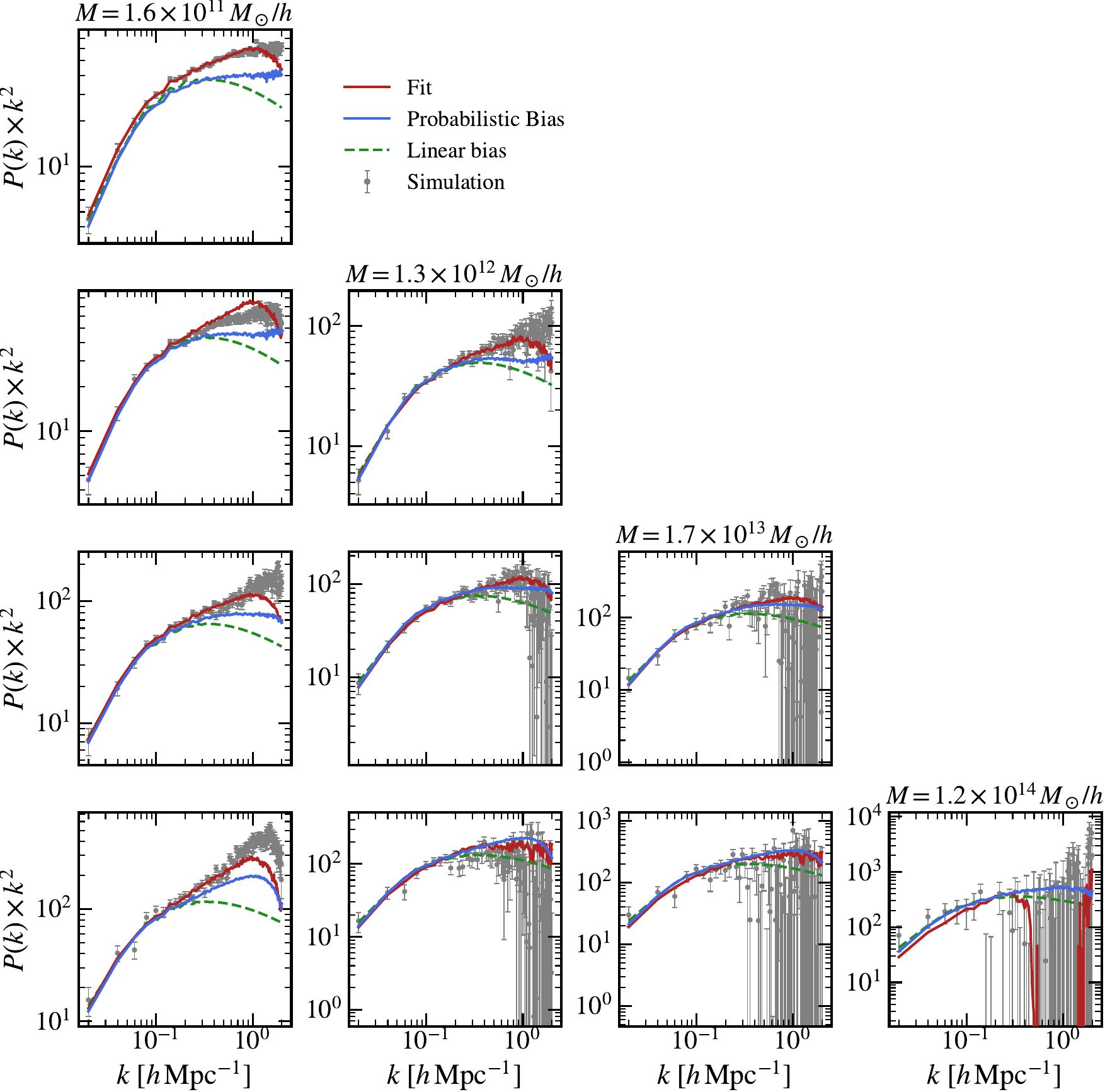}
\caption{Amplitude-level comparison between model predictions and simulation results for selected halo mass bins. We show $k^2P_{hh}(k|M_i,M_j)$ to make the scale dependence easier to inspect. Grey points show the simulation measurement; red lines show the best-fitting CHEFT model; blue lines show the CHEFT prediction using the probabilistic-bias coefficients; and green dashed lines show the standard linear-bias halo-model prediction $b^{\rm E}_1(M_i)b^{\rm E}_1(M_j)P_{\rm lin}(k)$. The fitted CHEFT model improves the description of the amplitudes of the auto- and cross-spectra relative to the probabilistic-bias and linear-bias predictions. However, matching these amplitudes does not guarantee an accurate description of the cross-mass coherence, shown in Fig.~\ref{fig:rhh_fit}.}
\label{fig:Phh_fit_amp}
\end{figure}

\begin{figure}
\includegraphics[width=0.98\columnwidth]{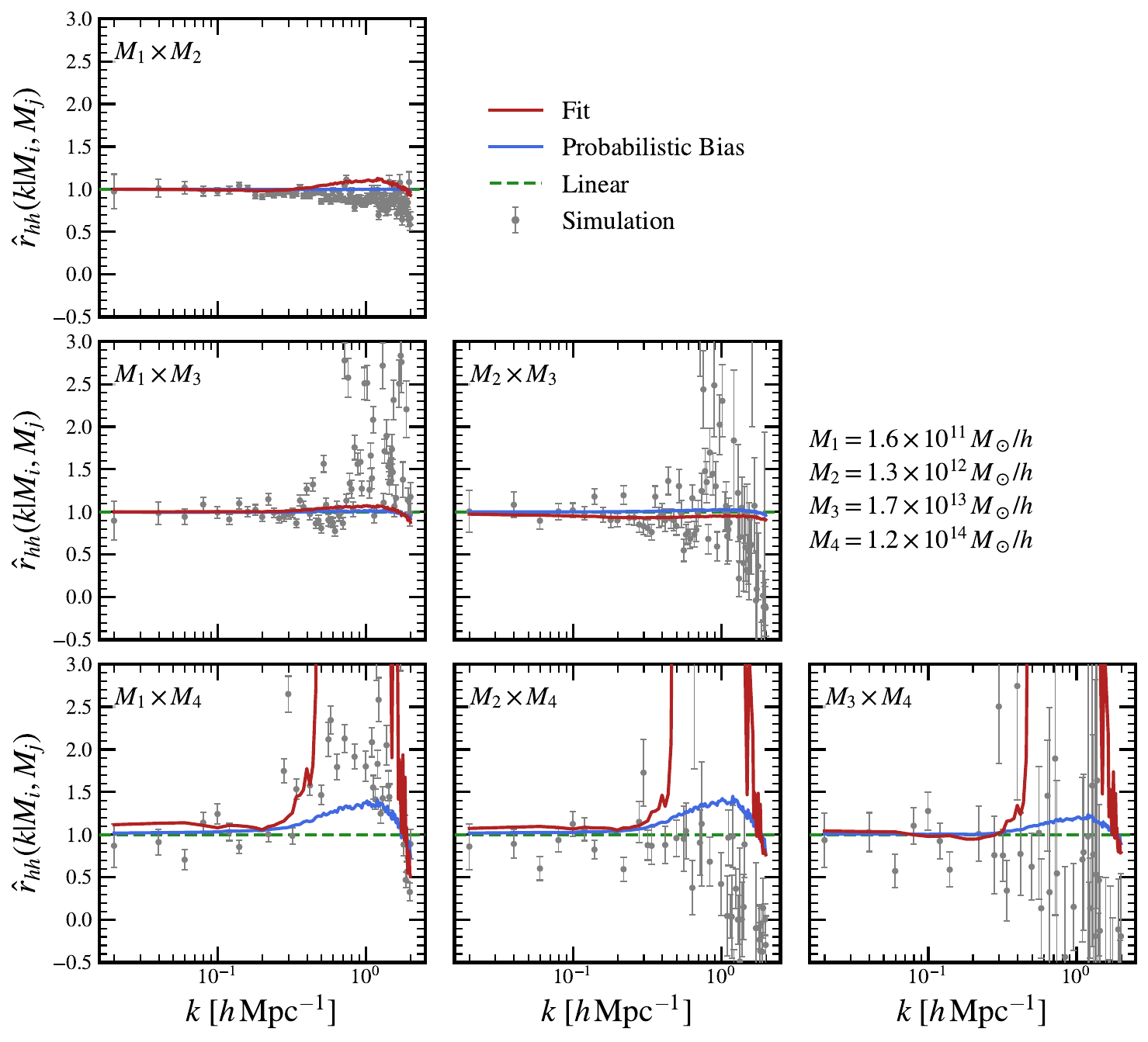}
\caption{Shot-noise-subtracted cross-mass coherence diagnostic for the same selected mass bins as in Fig.~\ref{fig:Phh_fit_amp}. We show $\hat r_{hh}^{\rm sub}(k|M_i,M_j)$, defined in Eq.~\eqref{eq:rhh_appendix}. Grey points show the simulation measurement; red lines show the best-fitting CHEFT model; blue lines show the CHEFT prediction using the probabilistic-bias coefficients; and green dashed lines show the standard deterministic linear-bias prediction. For this statistic, the linear-bias model corresponds to perfect coherence, $\hat r_{hh}^{\rm sub}=1$. Since the auto-spectra in the denominator have been shot-noise-subtracted, this quantity is not bounded by unity in noisy regimes; unphysical excursions above unity simply reflect sensitivity to stochasticity and shot-noise subtraction. The fitted CHEFT model captures part of the departure from the linear-bias expectation, but does not fully reproduce the scale dependence for the most widely separated mass bins.}
\label{fig:rhh_fit}
\end{figure}

This behaviour is illustrated in Figs~\ref{fig:Phh_fit_amp} and~\ref{fig:rhh_fit}. Figure~\ref{fig:Phh_fit_amp} shows examples of the measured spectra and model predictions for selected halo mass bins. We plot $k^2P_{hh}$ to make the scale dependence easier to inspect. The fitted CHEFT model reproduces the amplitudes of the auto-spectra and many of the cross-spectra more accurately than either the probabilistic-bias prediction or the standard linear-bias halo model. However, this amplitude-level agreement does not guarantee that the model captures the full off-diagonal structure of the halo--halo matrix.

To isolate this second aspect, Fig.~\ref{fig:rhh_fit} shows the corresponding
shot-noise-subtracted coherence diagnostic,
\begin{equation}
  \hat r_{hh}(k|M_1,M_2)  =   \frac{P^{\rm sub}_{hh}(k|M_1,M_2)}   {\left[P^{\rm sub}_{hh}(k|M_1,M_1)   P^{\rm sub}_{hh}(k|M_2,M_2)\right]^{1/2}}.
  \label{eq:rhh_appendix}
\end{equation}
Here $P^{\rm sub}_{hh}$ denotes the spectra after the shot-noise
subtraction used throughout this appendix. By construction, the standard deterministic linear-bias model predicts $r_{hh}=1$ for all mass pairs. The simulation results show departures from this simple expectation, especially for pairs involving widely separated halo masses. The fitted CHEFT model captures part of this behaviour, but does not fully reproduce the scale dependence of $r_{hh}$ in the most discrepant cases. This demonstrates that the residual error is associated with the cross-mass coherence structure of the halo field, rather than only with the clustering amplitude of individual mass bins.

\subsection{Interpretation}

These results show that the main limitation of the baseline CHEFT description is not simply an inability to assign the correct clustering amplitude to each halo mass bin. When the bias parameters are allowed to vary, the auto-spectra and many of the cross-spectrum amplitudes can be fitted accurately, as shown in Fig.~\ref{fig:Phh_fit_amp}. The difficulty appears when the same model is tested on the coherence between different halo populations, as shown in Fig.~\ref{fig:rhh_fit}. This indicates that the missing information is associated with the cross-mass coherence structure of the halo field.

In this sense, the failure of the full $P_{hh}(k|M_1,M_2)$ test should be interpreted as a limitation of the finite deterministic CHEFT basis used here. The model assumes that all halo populations can be represented as different linear combinations of the same collapsed Lagrangian operator fields. This is a significant generalisation of the standard fully halo mass factorised halo model, which forces $r_{hh}=1$, but it may still be insufficient if halo exclusion, stochasticity, or secondary halo properties introduce additional mass-dependent degrees of freedom. These effects can modify the off-diagonal entries of the halo-halo matrix without being fully determined by the auto-spectra, which explains why fitting $P_{hh}(k|M,M)$ does not guarantee a good prediction for $P_{hh}(k|M_1,M_2)$.

The preference for a model without additional noise terms and with a relatively small smoothing scale suggests that the dominant deterministic contribution to the clustering signal is already captured by the collapsed operator basis. At the same time, the residual discrepancies in $r_{hh}$ show that additional ingredients are required to describe the full mass dependence of halo clustering. We leave a systematic investigation of whether these degrees of freedom are best described by higher-derivative terms, halo exclusion, stochasticity, or assembly-dependent bias to future work.

\subsection{Calibration of the effective Laplacian model}
\label{ssec:laplacian_appendix}

In Section~\ref{ssec:non_mater.extended} we introduced the CHEFText model,
in which the fixed probabilistic-bias prediction for the Laplacian
bias function is replaced by the effective form
\begin{equation}
  b^{\rm eff}_{\nabla^2\delta}(M) =\alpha(M)b^{\mathcal{L},s}_{\nabla^2\delta}(M). \label{eq:beff_laplacian_appendix}
\end{equation}
Here $\smash{b^{\mathcal{L},s}_{\nabla^2\delta}(M)}$ is obtained by fitting a smooth
second-order polynomial to the probabilistic-bias measurements of
$b^L_{\nabla^2\delta}(M)$. This preliminary smoothing removes
small-scale fluctuations in the measured bias function before the
effective rescaling is applied.

We use the simple parametrisation
\begin{equation}
  \alpha(M)=c_0+c_1\log_{10}\left( \frac{M}{h^{-1}M_\odot}\right),
  \label{eq:alpha_laplacian_appendix}
\end{equation}
and determine $c_0$ and $c_1$ by fitting the model jointly to the
SZ-, CIB-, and HOD-like mock measurements, including both the
auto-spectra and the cross-spectra with matter. This gives
\begin{equation}
  c_0 = 1.036,  \qquad  c_1 = -0.072.
\end{equation}
The resulting effective bias function is shown as the orange dotted
curve in Fig.~\ref{fig:ProbBias}.

To test whether this correction is genuinely common to the different tracers, we also repeated the calibration separately for each weighting scheme. The resulting spectra changed only at the percent level relative to the joint calibration. This indicates that the effective Laplacian rescaling is not simply absorbing arbitrary tracer-specific freedom. Instead, within the tests performed here, it behaves as a common correction to the mass-dependent halo-halo clustering
model.

\begin{figure}
  \includegraphics[width=0.98\columnwidth]{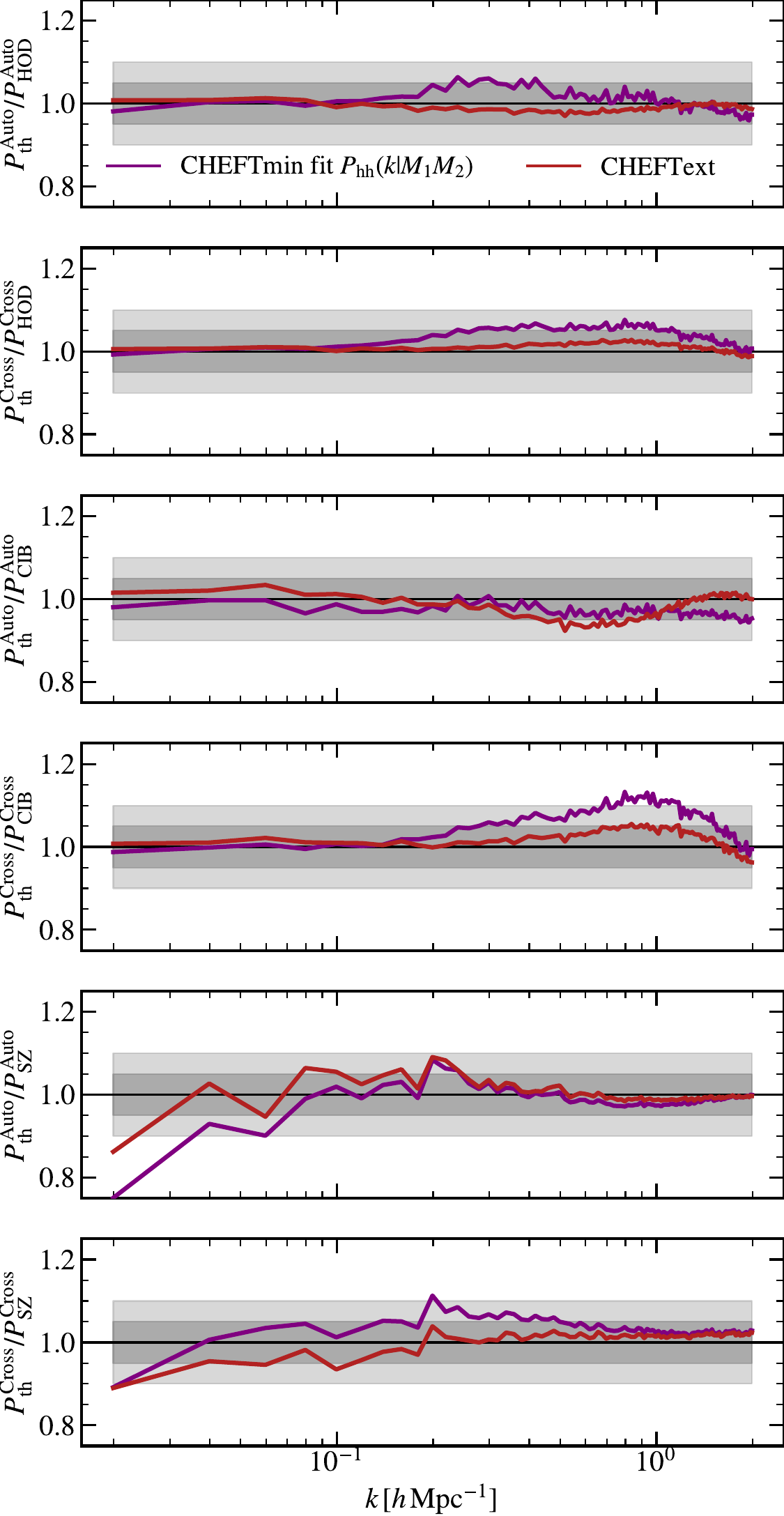}
  \caption{
  Comparison between the fully fitted CHEFT model and the effective Laplacian extension introduced in Section~\ref{ssec:non_mater.extended}. The fully fitted model uses bias parameters adjusted to the halo auto-spectra and halo-matter cross-spectra, while CHEFText uses a single effective Laplacian calibration applied to all tracers. Both approaches yield similar levels of accuracy for the reconstructed observables, with differences typically at the few-percent level. The largest residual difference appears in the CIB cross-spectrum. This supports the interpretation that the effective Laplacian rescaling captures the leading residual contribution without introducing separate tracer-dependent freedom.
  }
  \label{fig:spectra_corr}
\end{figure}

Figure~\ref{fig:spectra_corr} compares the predictions of CHEFText with those obtained from the more flexible model in which the CHEFT bias parameters are fitted directly to the halo auto-spectra and halo--matter cross-spectra. The two approaches yield similar levels of accuracy for the reconstructed observables, with differences typically below the few-percent level. The largest remaining difference appears in the CIB cross-spectrum. Given its simplicity and its apparent probe independence, we adopt the effective Laplacian model as the preferred extension of the hybrid framework in this proof-of-concept analysis.

\label{lastpage}
\end{document}